 \documentclass[twocolumn,showpacs,amsmath,prd,amssymb,showkeys,superscriptaddress,aps,longbibliography,nofootinbib]{revtex4-2}

\pdfoutput=1

\usepackage{graphicx}
\usepackage{dcolumn}
\usepackage{mathtools}
\usepackage{scalerel}
\usepackage{bm}
\usepackage{upgreek}
\usepackage{mathrsfs}
\usepackage{float}

  \usepackage{array}
 \usepackage{verbatim}
  \usepackage{amstext}
  \usepackage{amsmath}
  \usepackage{amssymb}
  \usepackage{slashed}
  \usepackage[bbgreekl]{mathbbol}
  \usepackage{indentfirst}
  \usepackage{xspace}
 \usepackage[T1]{fontenc}
 \usepackage[latin9]{inputenc}
 \usepackage{color}
 \PassOptionsToPackage{normalem}{ulem}
 \usepackage{ulem}
 \usepackage[unicode=true,pdfusetitle,
  bookmarks=true,bookmarksnumbered=false,bookmarksopen=false,
  breaklinks=false,pdfborder={0 0 1},backref=false,colorlinks=true]
  {hyperref}
 \hypersetup{
  citecolor=blue}

\newcommand{\nova}    {$NO$\nu$A$\xspace}

\makeatother

\begin{document}

\title{Search for Lorentz-violation through sidereal effect at NO$\nu$A Experiment}

\newcommand{\cusb}{Department of Physics,
Central University of South Bihar, Gaya 824236, India}

\newcommand{\bhu}{Department of Physics, Institute of Science,
Banaras Hindu University,
Varanasi 221005, India.}

\newcommand{\corrsks}{saurabhshukla@cusb.ac.in}
\newcommand{\corrls}{lakhwinder@cusb.ac.in}

\author{ Shashank~Mishra} \affiliation{ \cusb }\affiliation{ \bhu }
\author{ Saurabh~Shukla} \altaffiliation{ \corrsks }\affiliation{ \cusb }\affiliation{ \bhu }

\author{ Lakhwinder~Singh }  \altaffiliation{\corrls}\affiliation{ \cusb }
\author{ Venktesh~Singh }   \affiliation{ \cusb }

\date{\today}

\begin{abstract}

  Long-baseline neutrino oscillation experiments offer a unique laboratory to test the fundamental Lorentz symmetry, which is heart of both the standard model of particle and general relativity theory. The sidereal modulation in neutrino events will smoking-gun experimental signature of Lorentz and CPT violation. In this study, we investigate the impact of the sidereal effect on standard neutrino oscillation measurements within the context of the $\nova$ experiment.
  Additionally, we assess the sensitivity of the $\nova$ experiment to detect Lorentz-violating interactions, taking into account the sidereal effect. Furthermore, we highlight potential of the $\nova$ experiment to set the new constraints on anisotropic Lorentz-violating parameters.

\end{abstract}

\maketitle

\section{\label{sec:level1}Introduction:\protect}
Lorentz symmetry is a key assumption in our present understanding of high-energy processes and ensures the all inertial observers perceive the physical phenomenon identically. This symmetry, however, raises the question of testability in ultra-high energy theories at the Planck scale physics such as string theory~\cite{Kostelecky:1988zi, Kostelecky:1991ak}, loop quantum gravity~\cite{Gambini:1998it}, brane-worlds scenarios~\cite{CliffP.Burgess_2002}. These theories unify the gravity and gauge fields of the Standard Model (SM) of particle physics by allowing small perturbation of Lorentz symmetry, so-called Lorentz Invariance Violation (LIV)~\cite{Sahoo:2021dit}.  The Standard Model Extension (SME) serves as a effective theory of above mentioned ultra-high energy theories. The SME incorporates complete range of particles and interactions of SM as well as additional all possible  Lorentz violation operators, therefore, provides a feasible framework for LIV searches in a variety of scenarioss like gravity, charged leptons, photons, nucleons, and neutrinos~\cite{Colladay:1998fq, Bluhm:2005uj,RevModPhys.83.11}.

The discovery of ``finite neutrino masses and mixings'' with various neutrino sources is the first evidence of the existence of physics beyond the SM~\cite{Super-Kamiokande:1998qwk,KamLAND:2002uet,K2K:2006yov}. Over the last two decades, there is a tremendous development in neutrino experiments, allowing us  to enter the  era of precision measurement and the exploration of physics beyond the standard model. The neutrino sector therefore offers a novel venue to explore the LIV effect. LIV parameters are classified as isotropic and anisotropic. In experiments where both the neutrino source and detector are located on the Earth, the observed sidereal modulation in neutrino events provides the smoking-gun signature of a non-zero anisotropic LIV parameters. Several neutrino experiments have performed the analysis to study the LIV including MINOS-FD~\cite{MINOS:2008fnv}, MINOS-ND~\cite{MINOS:2010kat,MINOS:2012ozn}, IceCube~\cite{IceCube:2010fyu}, LSND~\cite{LSND:2005oop}, Super-K~\cite{Super-Kamiokande:2014exs}, T2K~\cite{T2K:2017ega}, DayaBay~\cite{DayaBay:2018fsh}, MiniBooNE~\cite{MiniBooNE:2011pix}, Double Chooz~\cite{DoubleChooz:2012eiq}, etc. Previous experimental searches for LIV using the sidereal effect have primarily concentrated on short baseline neutrino oscillation experiments and exclusively targeted the time-independent isotropic components~\cite{Barenboim:2018ctx}. The aim of this work is to expand and improve the sensitivity of LIV parameters in the non-isotropic time dependence case. We study the impact for sidereal time to the LIV parameters in the context of $\nova$ experiment which is a long baseline neutrino experiment~\cite{NOvA:2019cyt}. We study rotational symmetry for effective Hamiltonian which contain LIV and CPT perturbative terms. This restructured effective Hamiltonian  incorporates  into GLoBES software package to study the sensitivity of $\nova$ experiment towards non-isotropic LIV parameters in both appearance and disappearance channels.

This article is structured as follows. The general formulation of effective Hamiltonian
is discussed in Sec.~\ref{sect::formulation}. The effective Hamiltonian is also restructured to study sidereal effect. The main feature of this restructured effective Hamiltonian is discussed. Our approach to simualtion, adopted experimental design and standard oscillation parameters  are outlined in Sec.~\ref{sec:simulation}. In Sec.~\ref{sec::resDis}, we present sensitivity of $\nova$ experiment to explore LIV parameters in the presence of discrepancy of CP violating phase ($\delta_{cp}$) and $\theta_{23}$. Additionally, we also present upper limits for LIV parameters under sidereal analysis and
compare with  existing upper limits of LIV parameters from literature and conclude in Sec.~\ref{sec::summary}. 

\section{Formalism}
\label{sect::formulation}
In a lepton sector, the general form of the Lorentz violating part of the SME Lagrangian can be divided into CPT-even and CPT-odd terms \cite{Colladay:1998fq}: 
\setlength\abovedisplayskip{5pt}

\begin{equation}
  \begin{split}
    \mathcal{L}_{\rm{LIV}}^{CPT-even} & = -\frac{1}{2}(H_{L})_{\mu \nu A B} \bar{l}_{A}\sigma^{\mu \nu} {l}_{B}\\
    & + -\frac{1}{2}i(c_{L})_{\mu \nu A B} \bar{l}_{A}\gamma^{\mu} \leftrightarrow{D^{\nu}} {l_{B}}\\
    & + -\frac{1}{2}i(d_{L})_{\mu \nu A B} \bar{l}_{A}\gamma_{5}\gamma^{\mu} \leftrightarrow{D^{\nu}} {l}_{B},
    \label{lagrangain1}
  \end{split}
\end{equation}
where $(H_{L})_{\mu \nu A B}$ are antisymmetric coupling coefficients with dimensions of mass. $(c_{L})_{\mu \nu A B}$
and $(d_{L})_{\mu \nu A B}$ are symmetric and antisymmetric hermitian dimensionless CPT-even LIV coupling coefficients, respectively.

\begin{equation}
  \mathcal{L}_{\rm{LIV}}^{CPT-odd} = -(a_{L})_{\mu A B} \bar{l}_{A}\gamma^{\mu}{l}_{B} - (b_{L})_{\mu A B} \bar{l}_{A} \gamma_{5} \gamma^{\mu} {l}_{B},  
  \label{lagrangain2}
\end{equation}

where $(a_{L})_{\mu A B}$ and $(b_{L})_{\mu A B}$ are hermitian CPT-breaking LIV coupling coefficients  with dimension of mass.

In the Hamiltonian picture, the effective Hamiltonian ${(\mathcal{H}_{eff})}_{\alpha\beta}$ of neutrinos with small LIV and CPT violating perturbation is
generally written as~\cite{Diaz:2009qk}
\begin{equation}
  {(\mathcal{H}_{eff}})_{\alpha\beta} = (\mathcal{H}_{o})_{\alpha\beta} +  (\mathcal{H}_{LIV})_{\alpha\beta} ,    
  \label{TotHamil}
\end{equation}
where $(\mathcal{H}_{o})_{\alpha\beta}$ is a conventional standard neutrino Hamiltonian, describes the Lorentz-invariant neutrino oscillation and $ \mathcal({H}_{LIV})_{\alpha\beta}$ is a perturbative Hamiltonian including LIV contributions. The Indices~$\alpha$ and $\beta$ represent the three neutrino flavors. In general $({\mathcal{H}_{eff}})$ is a $6\times 6$ matrix which can be represented as:
\begin{equation}
  {(\mathcal{H}_{eff})} = 
  \left(
  \begin{array}{cc}
    (\mathcal{H}_{o})_{\nu\nu} &        0           \\
    0          & (\mathcal{H}_{o})_{\bar{\nu}   \bar{\nu}} 
  \end{array}
  \right)
  + \left(
  \begin{array}{cc}
    (\mathcal{H}_{liv})_{\nu\nu} & (\mathcal{H}_{liv})_{\nu \bar{\nu}} \\
    (\mathcal{H}_{liv})_{\bar{\nu}\nu}  & (\mathcal{H}_{liv})_{\bar{\nu}  \bar{\nu}} 
  \end{array}
  \right)
  \, ,
  \label{HamilMatrix}
\end{equation}
where $(\mathcal{H}_{o})_{\nu\nu}$($(\mathcal{H}_{o})_{\bar{\nu}  \bar{\nu}}$) is standard neutrino(anti-neutrino) Hamiltonian term which is responsible for standard neutrino(anti-neutrino) oscillations. Diagonal terms $(\mathcal{H}_{liv})_{\nu \nu}$ and $(\mathcal{H}_{liv})_{\bar{\nu} \bar{\nu}}$ are contributing for  neutrino-neutrino oscillation and antineutrino-antineutrino oscillation, respectively. Off-diagonal components, namely $(\mathcal{H}_{liv})_{\nu \bar{\nu}}$ and $(\mathcal{H}_{liv})_{\bar{\nu} \nu}$ govern neutrino-antineutrino oscillations and vice versa.

The standard neutrino(anti-neutrino) oscillation is parameterized by two mass square differences $\Delta m^{2}_{21}$, $\Delta m^{2}_{31}$, three mixing angles $\theta_{12}$, $ \theta_{23}$, $ \theta_{13}$ and a phase $\delta_{cp}$. In this study, we solely conform to neutrino-neutrino oscillation and corresponding Hamiltonian can be explicitly written as
\begin{equation}
  (\mathcal{H}_{o})_{\nu\nu} = {1\over{E}} \Big{[}  
 U \left(
  \begin{array}{ccc}
    0 &         0         &         0   \\
    0 & \Delta m^{2}_{21} &         0   \\ 
    0 &         0         & \Delta m^{2}_{31}
  \end{array}
  \right) U^{\dagger} + V_{matter}
 \Big{]} \, ,
  \label{HamilMatrix_2}
\end{equation}
where the PMNS matrix $U$ is parameterized as Ref ~\cite{Kopp:2007ne} and $V_{matter}$ is matter potential includes the matter effect. In the minimal SME, the interactions and neutrino propagation are both governed by the following leading-order effective hamiltonian ~\cite{Kostelecky:2003xn}
\begin{equation}
  ((\mathcal{H}_{liv})_{\nu\nu})_{\alpha\beta}  = |\vec{p}|\delta_{\alpha \beta} + \frac{1}{|\vec{p}|}[(a)^{\mu} p_{\mu}  - (c)^{\mu\nu} p_{\mu} p_{\nu}]_{\alpha \beta},
  \label{hLIV}
\end{equation}
where $(a)^{\mu}$ and $(c)^{\mu\nu}$ can be expressed as
\begin{equation}
  (a)^{\mu} = \frac{1}{2}( (a_{L}) + (b_{L}))^{\mu} ,
  (c)^{\mu\nu} = \frac{1}{2}( (c_{L}) + (d_{L}))^{\mu\nu}.  
  \label{aPlusB}
\end{equation}
$(a)^{\mu}$ and $(c)^{\mu\nu}$ are 3$\times$3 complex matrices represent LIV coefficients with mass dimension 1 and 0, respectively. The 4-momentum $p_{\mu}=(|\vec{p}|, \vec{p})$ introduces the energy and momentum dependencies in the Hamiltonian. It implies that the mixing behavior of neutrino flavor depends on the direction of neutrino propagation, which cause the rotational-symmetry violation. For the earth-based experiment, where the source and detector are fixed on the Earth's surface, the rotation of earth around its axis generates sidereal variation in oscillation probabilities. This variation can occur at multiples of the Earth's sidereal frequency $\omega_{\oplus}$ $\simeq$ 2$\pi$/(23 h 56 min). In order to comapre the results from different experiments, it is convenient to adopt a common inertial frame.  In the literature, measurements and sensitivities are conventionally expressed in terms of LIV coefficients defined in a Sun-centered celestial equatorial frame with coordinates (T,X,Y,Z).

The effective Hamiltonian with sidereal time dependencies in the Sun-centered celestial equatorial frame from Ref.~\cite{Kostelecky:2004hg} is restructured as follows:
\setlength\abovedisplayskip{5pt}
\begin{widetext}
\begin{equation}
  \begin{split}
       (\mathcal{H}_{liv})_{\alpha\beta} = (C)_{\alpha\beta} + R [ a_{\alpha\beta}^{X} - 2 E (c^{TX})_{\alpha\beta} + 2 E N_{z} (c^{XZ})_{\alpha\beta}] sin(\omega_{\oplus} T - \Phi_{orientation})~- \\
   R [ a_{\alpha\beta}^{Y} - 2 E (c^{TY})_{\alpha\beta} + 2 E N_{z} (c^{YZ})_{\alpha\beta}] cos(\omega_{\oplus} T - \Phi_{orientation})~+ \\
   R^{2} [ E \frac{1}{2}((c^{XX})_{\alpha\beta} - (c^{YY})_{\alpha\beta} ) ] cos(2 (\omega_{\oplus} T + \Phi_{orientation}))~+ \\
   R^{2} [ E (c^{XY})_{\alpha\beta}] sin(2 (\omega_{\oplus} T - \Phi_{orientation})),
  \end{split}
  \label{hLIVexpanded}
\end{equation}
\end{widetext}
where $T$ is the sidereal time, describes the earth's rotation w.r.t. a sidereal star in sun-centered frame.
Amplitude $(C)_{\alpha\beta}$, $\Phi_{orientation}$ and  $R$  can  be expressed  in the directional factors $N^{X}$, $N^{Y}$, $N^{Z}$ in the following manner: 
\begin{equation}
  \Phi_{orientation} = \tan^{-1}(N^{Y}/N^{X}),
\end{equation}
\begin{equation}
  R = \sqrt{ N_{X}^2 + N_{Y}^2 }, 
\end{equation}
\begin{equation}
  \begin{split}
(C)_{\alpha\beta} &= (a)^{T}_{\alpha \beta} - N^{Z}(a)^{Z}_{\alpha \beta} + E[-\frac{1}{2}(3 - N^{Z} N^{Z})(c)^{TT}_{\alpha \beta} \\
&+ 2N^{Z}(c)^{TZ}_{\alpha \beta} + \frac{1}{2}(3 - N^{Z} N^{Z})(c)^{ZZ}_{\alpha \beta}].
  \end{split}
\end{equation}
The directional factors ($N^{X}$, $N^{Y}$, $N^{Z}$) are further expressed in terms of the angle between the beam and the vertically upward direction ($\theta$) known as ``Zenith'' angle; the angle between the beam and the south measured towards the east ($\phi$) known as ``bearing'' angle; and the colatitude of the detector ($\chi$)~\cite{Kostelecky:2004hg}.
\begin{equation}
  \begin{split}
  N^{X} &= \cos\chi\sin\theta\cos\phi + \sin\chi\cos\theta,\\
  N^{Y} &= \sin\theta\sin\phi,\\
  N^{Z} &= -\sin\chi\sin\theta\cos\phi + \cos\chi\cos\theta, 
  \end{split}
  \label{orientationbeam}
\end{equation}
 The LIV coefficients $(a)_{\alpha \beta}^{\mu}$ are solely governed by the baseline, while  coefficients $(c)_{\alpha\beta}^{\mu\nu}$ are subject to control from both the baseline length and the energy of the neutrinos.
 The parameters $(a)^{T}_{\alpha \beta}$,$(a)^{Z}_{\alpha \beta}$,  $(c)^{TT}_{\alpha \beta}$,  $(c)^{TZ}_{\alpha \beta}$ and $(c)^{ZZ}_{\alpha \beta}$ belong to $(C)_{\alpha\beta}$ has no sidereal time dependency in the perturbation, while the parameters $(a)^{X}_{\alpha \beta}$, $(a)^{Y}_{\alpha \beta}$,  $(c)^{TX}_{\alpha \beta}$,  $(c)^{TY}_{\alpha \beta}$, $(c)^{XX}_{\alpha \beta}$,  $(c)^{XY}_{\alpha \beta}$,  $(c)^{XZ}_{\alpha \beta}$, $(c)^{YY}_{\alpha \beta}$ and $(c)^{TZ}_{\alpha \beta}$ are responsible for sidereal modulation of perturbed Hamiltonian terms. Lorentz violation manifests exclusively in particle transformations rather than observable transformations. It implies the perturbation terms remain invariant in a rotation transformation of frame in $XY$ plane. When we apply the transformation   $\phi_{orientation} \rightarrow \phi_{orientation} + \pi/2$, the following interrelations among the LIV parameters emerge:

\begin{equation}
  \begin{split}
    ({a^{X}})_{\alpha \beta}  & \rightarrow  ({a^{Y}})_{\alpha \beta},\\
    ({a^{Y}})_{\alpha \beta}  & \rightarrow  -({a^{X}})_{\alpha \beta}, \\
    ({c^{TX}})_{\alpha \beta} & \rightarrow ({c^{TY}})_{\alpha \beta} ,\\
    ({c^{TY}})_{\alpha \beta} & \rightarrow -({c^{TX}})_{\alpha \beta}, \\
    ({c^{XZ}})_{\alpha \beta} & \rightarrow ({c^{YZ}})_{\alpha \beta} ,\\
    ({c^{YZ}})_{\alpha \beta} & \rightarrow -({c^{XZ}})_{\alpha \beta}, \\
     ({c^{XY}})_{\alpha \beta} & \rightarrow -({c^{XY}})_{\alpha \beta}, \\
    ({c^{XX})_{\alpha \beta} - (c^{YY}})_{\alpha \beta} & \rightarrow ({c^{XX})_{\alpha \beta} - (c^{YY}})_{\alpha \beta},
  \end{split}
  \label{eq15}
\end{equation}

The invariance of observable transformations leads to a reduction in the number of relevant LIV parameters from 27 to 12. Consequently, this study focuses exclusively on 12 parameters, which include $(a)^{X}_{e\mu}$, $(a)^{X}_{e\tau}$, $(a)^{X}_{\mu\tau}$, $(c)^{TX}_{e\mu}$, $(c)^{TX}_{e\tau}$, $(c)^{TX}_{\mu\tau}$, $(c)^{XX}_{e\mu}$, $(c)^{XX}_{e\tau}$, $(c)^{XX}_{\mu\tau}$, $(c)^{XZ}_{e\mu}$, $(c)^{XZ}_{e\tau}$, $(c)^{XZ}_{\mu\tau}$.

If the contribution of LIV perturbation in Eq.~\ref{TotHamil} is sufficiently small, the oscillation probabilities for both the appearance and disappearance channels can be expressed up to the leading order for the $\mu e$ and $\mu\mu$ channels, similarly as presented in Ref~\cite{Liao:2016hsa, Chaves:2018sih, Dey:2018yht,Yasuda:2007jp, Masud:2015xva, Masud:2016bvp, Masud:2016gcl, Majhi:2019tfi}.
\begin{widetext}
  \begin{equation}
  \begin{split}
     P_{\mu e}^{\rm LIV}  \simeq x^2 f^2 +2 x y fg \cos (\Delta+\delta_{CP})
    + y^2 g^2 + 4 r_A |{h}^{\rm{LIV}}_{e \mu}| 
    \big\{ xf \big [ f s_{23}^2 \cos (\phi^{\rm{LIV}}_{e \mu}+\delta_{CP})
      +g c_{23}^2 \cos( \Delta +\delta_{CP}+\phi^{\rm{LIV}}_{e \mu})\big] \\
     + yg \big[ g c_{23}^2 \cos \phi^{\rm{LIV}}_{e \mu} 
      + f s_{23}^2\cos (\Delta-\phi^{\rm{LIV}}_{e \mu})\big ]\big \}
    + 4 r_A |{h}^{\rm{LIV}}_{e\tau}| s_{23} c_{23} \big\{xf \big[ f \cos (\phi^{\rm{LIV}}_{e \tau}+\delta_{CP})
      - g\cos (\Delta +\delta_{CP}+\phi^{\rm{LIV}}_{e \tau}) \big] \\
       - yg[ g \cos \phi^{\rm{LIV}}_{e \tau}
      -f \cos(\Delta-\phi^{\rm{LIV}}_{e\tau})\big]\big\}+ 4 r_A^2 g^2 c_{23}^2 | c_{23} |{h}^{\rm{LIV}}_{e \mu}|
    -s_{23} |{h}^{\rm{LIV}}_{e\tau}||^2 + 4 r_A^2 f^2 s_{23}^2  | s_{23} |{h}^{\rm{LIV}}_{e \mu}|
    +c_{23} |{h}^{\rm{LIV}}_{e\tau}||^2 \\
     + 8 r_A^2 f g s_{23}c_{23} \big\{ c_{23} \cos \Delta \big[s_{23}(|{h}^{\rm{LIV}}_{e \mu}|^2 -|{h}^{\rm{LIV}}_{e \tau}|^2)
      + 2 c_{23} |{h}^{\rm{LIV}}_{e \mu}| |{h}^{\rm{LIV}}_{e \tau}|\cos(\phi^{\rm{LIV}}_{e \mu}-\phi^{\rm{LIV}}_{e \tau}) \big] \\
     - |{h}^{\rm{LIV}}_{e \mu}|| {h}^{\rm{LIV}}_{e \tau}| \cos (\Delta -\phi^{\rm{LIV}}_{e \mu}
    +\phi^{\rm{LIV}}_{e \tau})\big \} 
    +{\cal O}(s_{13}^2 a, s_{13}a^2, a^3),
    \label{pmue}
\end{split}
\end{equation}

\begin{equation}
  \begin{split}
    P_{\mu\mu}^{\rm LIV} & \simeq 1- \sin^2 2 \theta_{23}\sin ^2 \Delta  - |{h}^{\rm{LIV}}_{\mu\tau}|
    \cos \phi^{\rm{LIV}}_{{\mu\tau}} \sin 2 \theta_{23}
    \Big[ (2r_A\Delta )\sin^2 2\theta_{23}\sin 2\Delta + 4 \cos^2 2 \theta_{23}r_A\sin ^2\Delta\Big]\\
    & + (|{h}^{\rm{LIV}}_{\mu\mu}| - |{h}^{\rm{LIV}}_{\tau\tau}|)\sin^2  2 \theta_{23} \cos 2 \theta_{23}
    \Big[(r_A\Delta) \sin 2\Delta -2r_A \sin ^2 \Delta  \Big],
    \label{pmumu}
  \end{split}
\end{equation}
where
\begin{equation}
  \begin{split}
    s_{ij}=\sin\theta_{ij},~~c_{ij}=\cos\theta_{ij},~~ 
    x=2s_{13}s_{23},~~ y=2rs_{12}c_{12}c_{23},~~  r=|\Delta m^2_{21}/\Delta m^2_{31}|,~~
     \Delta = \frac{\Delta m^2_{31} L}{4E},~~   \\
     V_{CC}=\sqrt 2 G_F N_e,~~ 
     r_A=\frac{2E}{{\Delta m}^2_{31}},~~
    f=\frac{\sin\big[\Delta (1-r_A(V_{CC}+{h}^{\rm{LIV}}_{ee}))  \big]}{1-r_A(V_{CC}+{h}^{\rm{LIV}}_{ee})},~~
    g=\frac{\sin\big[\Delta r_A(V_{CC}+{h}^{\rm{LIV}}_{ee})  \big]}{r_A(V_{CC}+{h}^{\rm{LIV}}_{ee})}.\\
    \hspace{0.5 true cm}\label{os-po}
  \end{split}
\end{equation}
\end{widetext}

  The antineutrino probability $ P_{\bar \mu \bar e}^{\rm LIV}$ ($P^{LIV}_{\bar{\mu} \bar{\mu}}$) can be obtained from Eq.~\ref{pmue}(Eq.~\ref{pmumu}) by replacing $V_{CC} \to -V_{CC}$, $\delta_{CP} \to - \delta_{CP}$ and 
$a_{\alpha \beta} \to -a_{\alpha \beta}^*$. Similar expression for inverse hierarchy can be obtained by substituting $\Delta m_{31}^2 \to - \Delta m_{31}^2$, i,e., $\Delta \to -\Delta$ and $r_A \to - r_A$.

  \section{Numerical Procedure of simulation}
  \label{sec:simulation}
  NuMI Off-Axis $\nu_{e}$ Appearance Experiment ($\nova$), a long baseline experiment at Fermilab, examines neutrino oscillations using a high-intensity and high-purity beam of either muon neutrinos or muon antineutrinos. The experiment utilizes two identical detectors: a Far Detector (FD) and a Near Detector (ND). The fiducial mass of FD is 14 kTon, and it is situated 810 KM away from the target and 14 mRad off axis~\cite{NOvA:2007rmc}. As a fixed baseline experiment, $\nova$ can observe the sidereal variation in the neutrino event rate in FD arising from the Earth's rotation.  In order to study the oscillation probabilities and event rate for $\nova$ experiment, we adopted GLoBES~\cite{Huber:2004ka}~\cite{Huber:2007ji} software package with suitable modifications in $snu.c$ plugin to include the sidereal effect. A exposure total of $2.5\times 10^{21}$ protons on target (POT) is utilized for the analysis of neutrinos, and an identical exposure is applied for antineutrinos. The POT is independent of sidereal time and remains constant throughout the time bin. For both the appearance and disappearance channels, the energy window is fixed from 1.0 GeV to 5.0 GeV, with a peak value at 2.0 GeV.
\begin{table}[!ht]
  \caption{The standard oscillation parameters are  used in this work~\cite{Esteban:2020cvm}.}
  \label{table2OScPar}
     \setlength{\tabcolsep}{14pt}
 \renewcommand{\arraystretch}{1.5}
  \begin{tabular}{|c|c|c|}
    \hline 
    Parameter            & True Value      &  Test Value\\ \hline \hline
    $\theta_{12}$         & $33.48^{\circ}$ &  --\\ \hline
    $ \theta_{13}$        & $8.5^{\circ}$   &  --\\ \hline
    $ \theta_{23}$        & $45.0^{\circ}$  &  $(41.0^{\circ},52.0^{\circ}) $\\ \hline
    $\delta_{cp}$         &$ 195.0^{\circ}$  &  $(0^{\circ},360.0^{\circ}) $ \\ \hline
    $\triangle m^{2}_{21}$    & $7.55 \times 10^{-5} eV^{2}$  & --\\ \hline 
    $\triangle m^{2}_{31}$    & $2.50  \times 10^{-3} eV^{2}$  & --\\ \hline
     \end{tabular}
\end{table}
\begin{table}[!ht]
  \caption{$\nova$ FD orientation details used in the simulation~\cite{NOvA:2007rmc}.}
  \label{table3beamO}
   \setlength{\tabcolsep}{15pt}
 \renewcommand{\arraystretch}{1.5}
  \begin{tabular}{|c|c|}
    \hline 
    Parameter            & Value         \\ \hline \hline
    $\chi$   co-latitude  & $ 48.3793^{\circ} $\\   \hline
    $\theta$ zenith angle   & $ 84.26^{\circ} $\\   \hline
    $\phi $  bearing       & $204.616^{\circ} $\\   \hline
      \end{tabular}
\end{table}

Table~\ref{table2OScPar} provides a summary of the standard oscillation parameters used in this work. 
Since $\nova$ is not sensitive for the mixing angle $\theta_{12}$ and $\theta_{13}$~\cite{denton2023comes},and these parameters  are well-measured by other neutrino oscillation experiments,  hence their values is fixed in simulation.
The latest data of  $\nova$ experiment favor the normal neutrino mass hierarchy by 1.9$\sigma$ \cite{NOvA:2019cyt}, therefore the normal mass ordering is also fixed throughout the simulation. Details on the beam orientation and FD of $\nova$ experiment, which is employed for the simulation, are represented in Table~\ref{table3beamO}. 

\begin{figure}[!ht]
  \centering
  \includegraphics[height= 0.40\textwidth,width=0.50\textwidth]{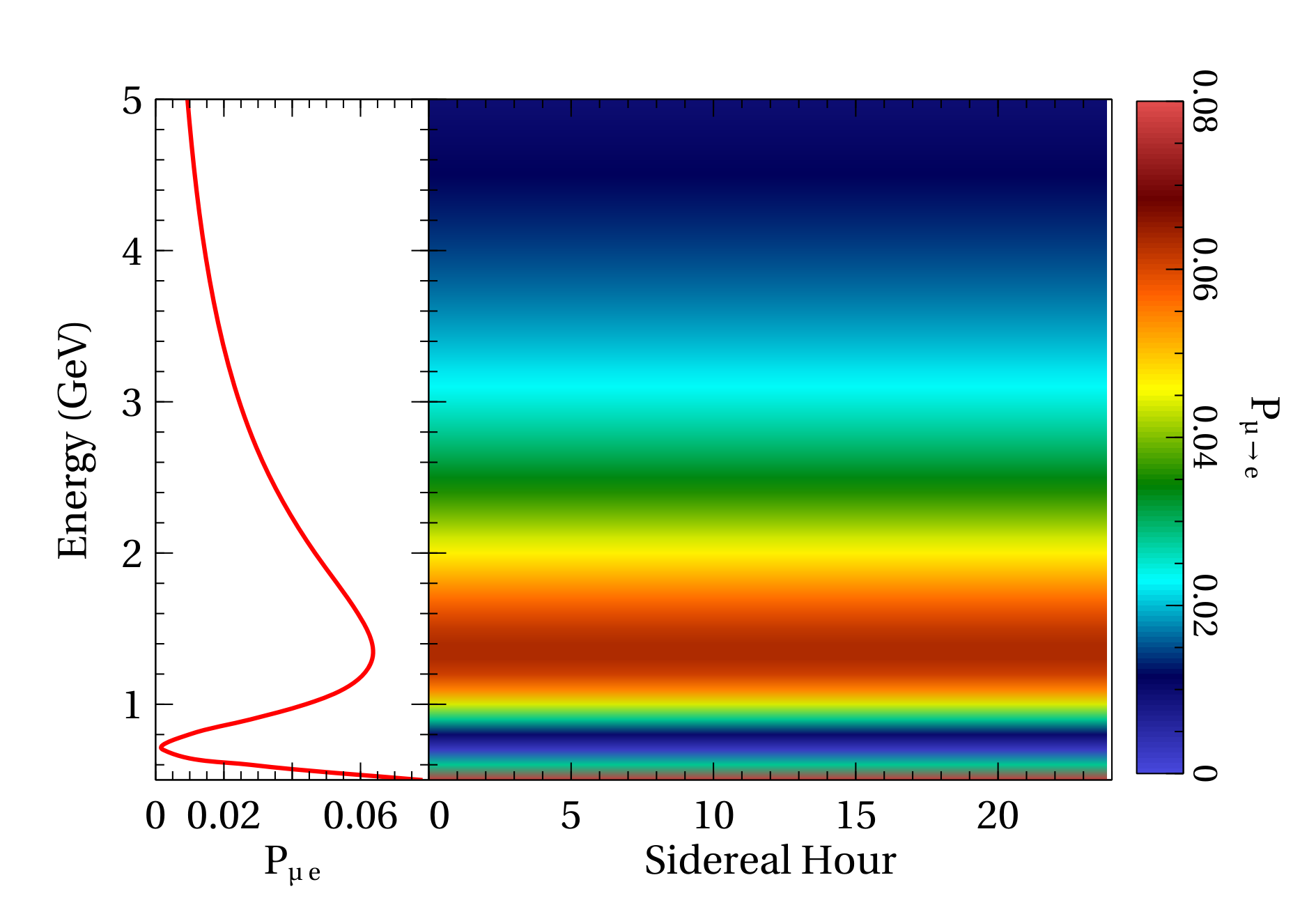}
  \includegraphics[height= 0.40\textwidth,width=0.50\textwidth]{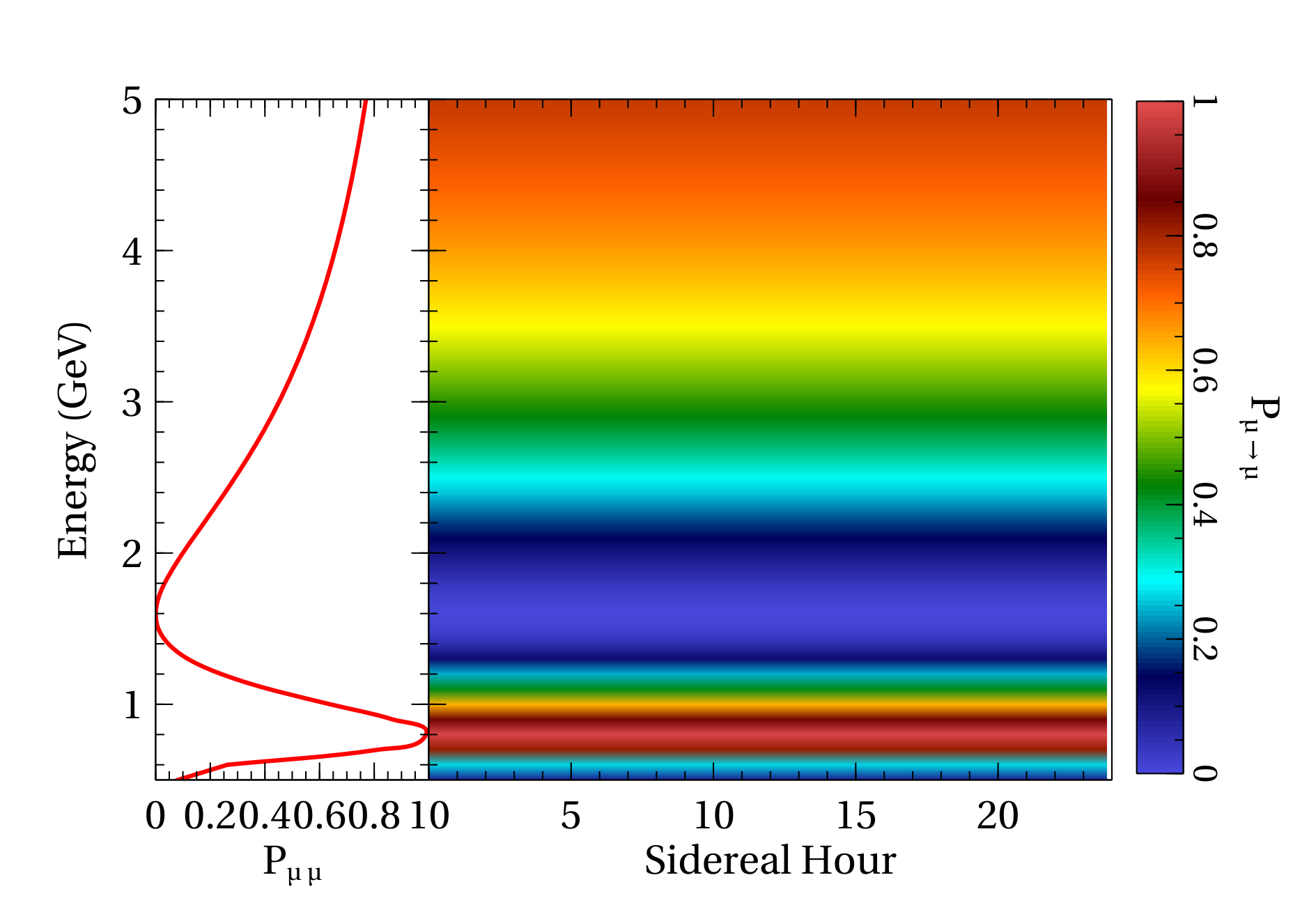}
  \caption{The standard neutrino oscillation 1-D probability spectrum in terms of energy, as well as the probability distribution in terms of local sidereal time (LST) and energy for the appearance channel (top) and disappearance  channel (bottom)  without taking LIV parameters into account. The oscillation parameters listed in Table~\ref{table2OScPar} is adopted to calculate the probability distribution. }
 \label{Prob2DS} 
\end{figure}

\begin{widetext}
  \begin{minipage}{\linewidth}
    \begin{figure}[H]
      \centering
   \includegraphics[height= 0.55\textwidth,width=0.90\textwidth]{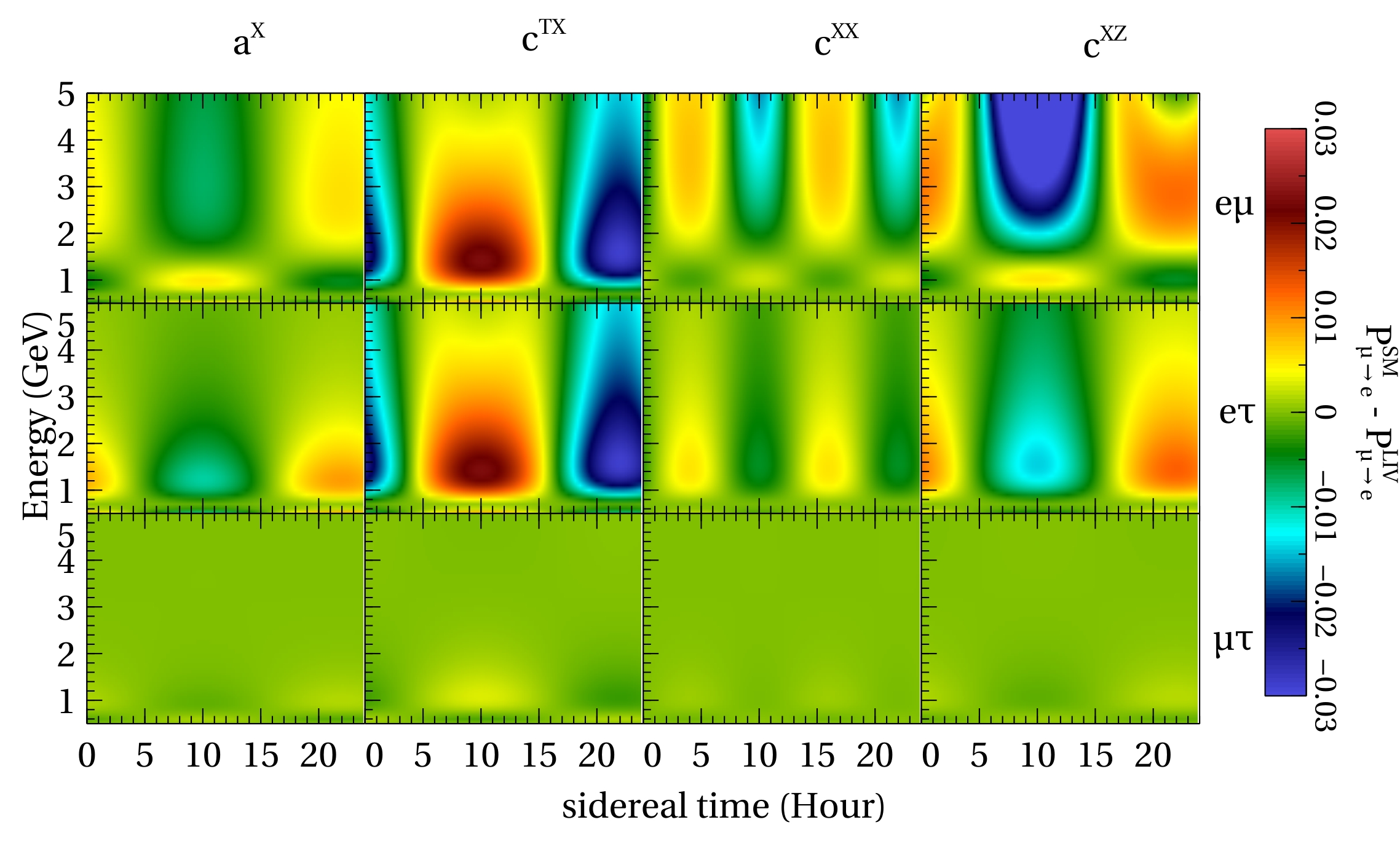}
   \caption{The probability distribution for the appearance channel  is depicted in each panel corresponding to a specific non-zero LIV parameter.  In each panel, one specific LIV parameter is set to $1\times 10^{-23}$, while the others are set to 0.} 
   \label{Fig::Prob2DLIVAppCh}
 \end{figure}
 \end{minipage}
  \begin{minipage}{\linewidth}
    \begin{figure}[H]
       \centering
 \includegraphics[height= 0.55\textwidth,width=0.90\textwidth]{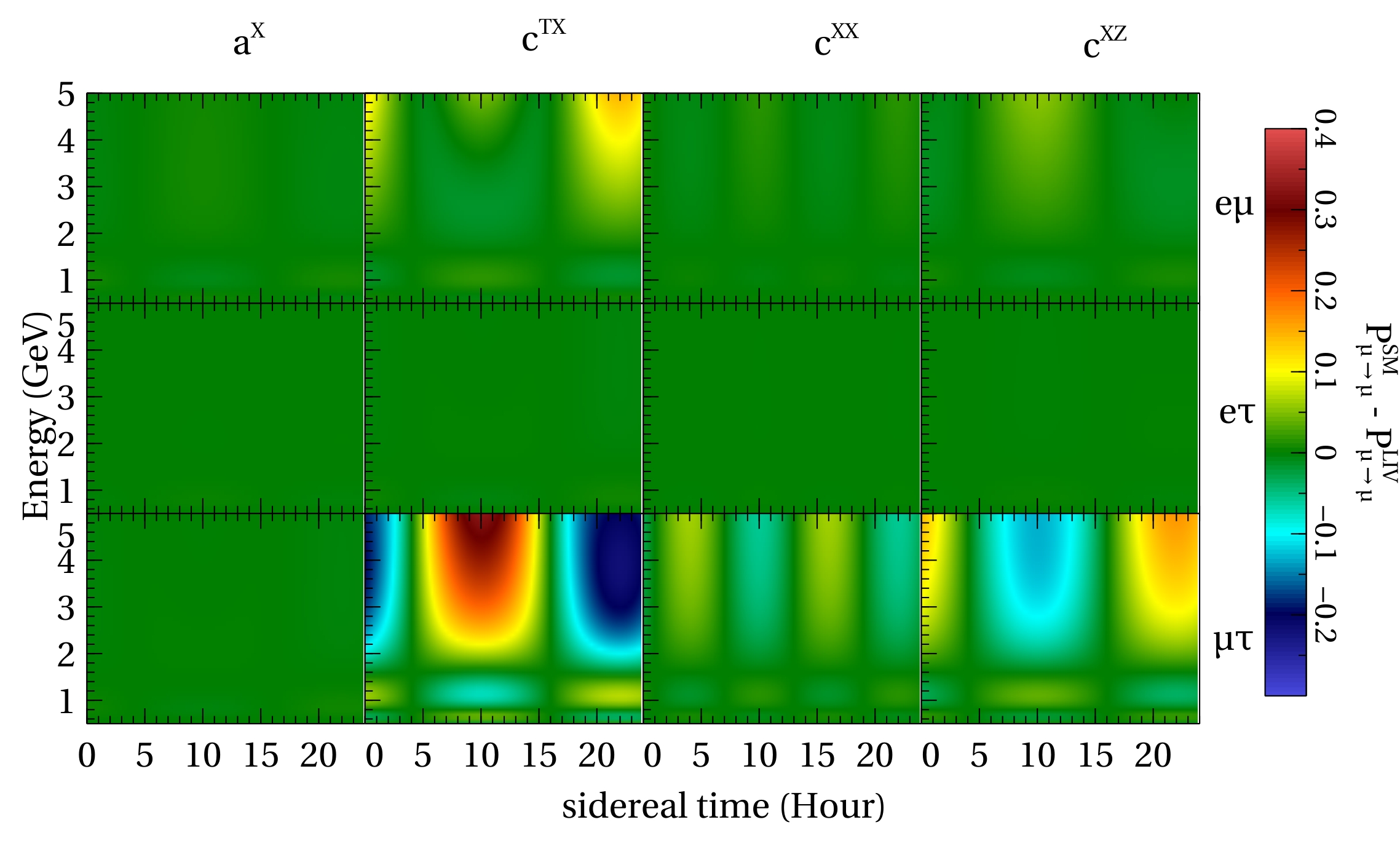}  
 \caption{The probability distribution for the disappearance channel  is depicted in each panel corresponding to a specific non-zero LIV parameter.  In each panel, one specific LIV parameter is set to $1\times 10^{-23}$, while the others are set to 0.}
\label{Fig::Prob2DLIVDisCh} 
 \end{figure}
 \end{minipage}
 \end{widetext}
\section{Results and Discussion}
\label{sec::resDis}
The standard neutrino oscillation probability spectrum without LIV parameters for appearance  and disappearance  channels with respect to energy and local sidereal time (LST) is depicted in top and bottom of the Fig.~\ref{Prob2DS}, respectively. All energies of neutrinos have a smooth probability distribution throughout the whole sidereal time. However, there is a considerable distortion in the standard neutrino oscillation probability distribution when LIV parameters are taken into account. In order to analysis the impact of LIV parameters on the probability-LST distribution, a series of tests is designed with one LIV coefficient set to be very  small value such as $1\times 10^{-23}$ and all other LIV coefficients set to zero.  The probability distribution difference between SM and LIV for the appearance and disappearance channel is shown in Figs.~\ref{Fig::Prob2DLIVAppCh} and~\ref{Fig::Prob2DLIVDisCh}, respectively. The first, second, third, and fourth panels (from left to right) of Figs.~\ref{Fig::Prob2DLIVAppCh} and~\ref{Fig::Prob2DLIVDisCh} illustrate the distortion in the standard neutrino oscillation probability distribution, when $(a^{X})_{\alpha \beta}$, $(c^{XX})_{\alpha \beta}$, $(c^{TX})_{\alpha \beta}$, and $(c^{XZ})_{\alpha \beta}$ parameters set to non-zero value, respectively. The top, middle, and bottom panels represent ${\alpha \beta}$ = $e\mu$, $e\tau$, and  $\mu\tau$ coefficients, respectively.

As we move from left to right in the top panel of Figs.~\ref{Fig::Prob2DLIVAppCh}~and~\ref{Fig::Prob2DLIVDisCh}, the probability  modulates as $\omega_{\oplus}$, $\omega_{\oplus}$, 2$\omega_{\oplus}$ and $\omega_{\oplus}$. This nature is fairly evident for $e\mu$ and $e\tau$ coefficients in the appearance channel as well as $\mu\tau$ coefficients in the disappearance channel, such modulations in  data serve as somking-gun signature for LIV. This particular feature can also be comprehended by referring to Eq.~\ref{hLIVexpanded}, which demonstrates the relative contribution of each parameter and dependency on LST.

In appearance channel,there is an enhancement in the sidereal modulation probability near 2 GeV due to $(c)^{TX}_{e\mu}$, $(c)^{TX}_{e\tau}$, and $(c)^{XZ}_{e\tau}$. Conversely, the modulation strength for $(a)^{X}_{e\mu}$, $(c)^{XX}_{e\mu}$, and $(c)^{XZ}_{e\mu}$ increases in the region of 4 GeV. At the leading order, there is no involvement of LIV parameters associated with the $\mu \tau$ coefficients in sidereal modulations.  The LIV parameters associated with the $e \mu$ and  $e \tau$ coefficients  do not play a significant role in the disappearance channel. This is evident as they do not appear in the leading-order term of the disappearance channel probability. However, for LIV parameters $(c)^{TX}_{\mu \tau}$, $(c)^{XX}_{\mu \tau}$ and $(c)^{XX}_{\mu \tau}$ have significantly larger amplitude in  the modulation around  4 GeV energy region. 
  
  The primary objectives of all ongoing and prospective high-precision long-baseline neutrino oscillation experiments are to determine the precise CP violating phase ($\delta_{cp}$) and the octant of $\theta_{23}$, as well as resolving the mass hierarchy. However, there is significant uncertainties in the current measurement of $\theta_{23}$ and $\delta_{cp}$ phase. In the case of long-baseline searches, standard oscillation parameters  mix with  LIV parameter. The  unknown standard oscillation parameters ($\delta_{cp}$, $\theta_{23}$) introduce a level of uncertainty that can potentially reduce the sensitivity of the experiment to detect the sidereal signal. We therefore  investigate the correlations between the LIV parameters and conventional oscillation parameters $\theta_{23}$  and $\delta_{cp}$. We also assess  sensitivity of the $\nova$ experiment to the sidereal effect.

\subsection{\label{sec:citeref}Sensitivity}
In order to derive the sensitivity, we adopt the Poisson-likelihood chi-square statistics. The Poisson-likelihood chi-square function for $\nova$ experiment can be  written as:~\cite{Baker:1983tu}
\setlength\abovedisplayskip{5pt}

\begin{equation}
  \begin{split}
    &\chi^{2}_{total}( N_{test},  N_{true})  = \\
    &\sum_{\substack{i,j,k}} 2 \left( N_{test}^{i j k} - N_{true}^{i j k} + N_{true}^{i j k} \times ln\left[\frac{ N_{true}^{i j k}}{ N_{test}^{i j k} }\right] \right), 
  \end{split}
  \label{chisq}
\end{equation}
where "i" stands for LST bins, "j" for appearance and disappearance channels, and "k" for the beam's neutrino and anti-neutrino modes. The "$N_{true}$" represents the total number of events in each sidereal bin for energy window of 1 to 5 GeV in the SM case, while "$N_{test}$" represents the same quantity in the case of LIV. We adopt 24 sidereal bins, each spanning one sidereal hour, covering the entire duration of a sidereal day. The total 5\% of systematic uncertainty is considered in final analysis. Systematics is incorporated using so called pull method.

The strength of a LIV parameter depends on its phase, therefore, the sensitivity of experiment towards particular LIV parameter is influenced by the phase of that parameter. As the phases of these parameters are unknown, we perform the marginalization over full parameter space of LIV phase ($\phi_{\textrm{parameter}}$) along with uncertainty range of $\delta_{CP}$ to investigate the correlation between  LIV parameters and $\theta_{23}$. Figure~\ref{corelation_th} illustrates the correlation between $\delta_{CP}$ and  non-diagonal LIV parameters ($a^{X}_{\alpha\beta}$, $c^{TX}_{\alpha\beta}$, $c^{XX}_{\alpha\beta}$, $c^{XZ}_{\alpha\beta}$ with  ${\alpha \beta}$ = $e\mu$, $e\tau$, and  $\mu\tau$)  at $2\sigma$, $2.5\sigma$, and $3\sigma$ significance level.
The contribution of LIV parameters corresponding to $e \tau $ coefficient for appearance channel in oscillation probability is suppressed by approximately factor of 2, due to $sin(\theta_{23})cos(\theta_{23})$ term.
Therefore, sensitivity of these parameters degrade as compare to  LIV parameters corresponding to $e \mu $ coefficient for appearance channel. Figure~\ref{corelation_cp} shows correlation between  $\delta_{CP}$ phase and non-diagonal LIV parameters at  $2\sigma$, $2.5\sigma$, and $3\sigma$ significance level with marginalizing over the both $\phi_{\textrm{parameter}}$ and $\theta_{23}$. To assess the sensitivity of $\nova$ to the non-diagonal  LIV parameters and across the entire range of corresponding phase values, we marginalize over $\theta_{23}$ and $\delta_{CP}$ phase.  Figure \ref{corelation_ph} illustrates the allow region of non-diagonal LIV parameters with respect to the entire range of corresponding $\phi_{\textrm{parameter}}$ at $2\sigma$, $2.5\sigma$, and $3\sigma$ significance level. One can notice that the sensitivity  of non-diagonal LIV parameters corresponding $\mu \tau$ disappearance channel is enhanced when  $\phi_{\textrm{parameter}}$ is purely real as compare to purely imaginary.  The non-diagonal LIV parameters corresponding to appearance channel do not have such feature due to marginalization over $\theta_{23}$ and $\delta_{CP}$ phase.

\begin{widetext}
     \begin{minipage}{\linewidth}
    \begin{figure}[H]
    \includegraphics[height= 0.50\textwidth,width=0.9\textwidth]{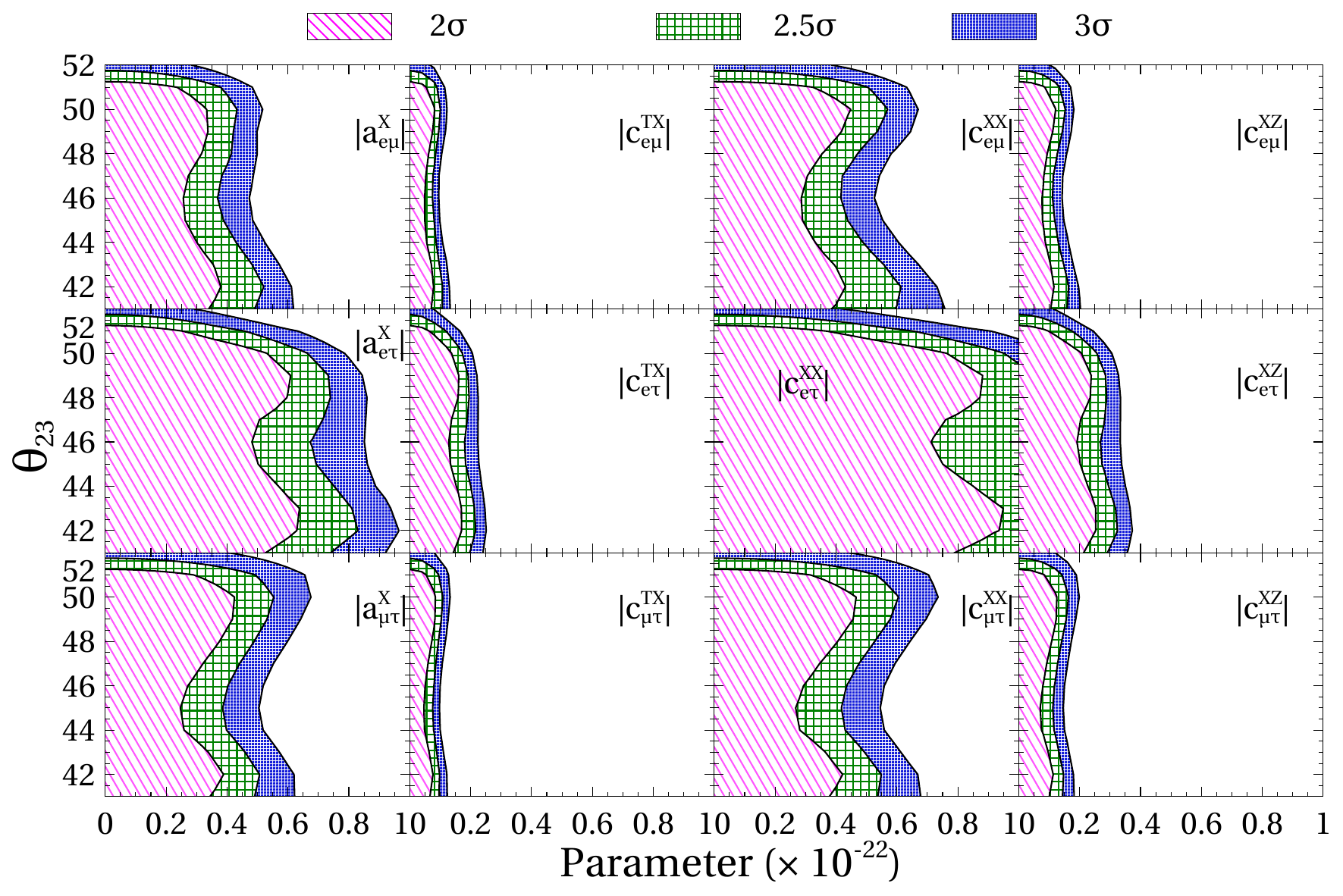}
    \caption{Correlations between the non-diagonal parameters ($a^{X}_{\alpha\beta}$, $c^{TX}_{\alpha\beta}$, $c^{XX}_{\alpha\beta}$, $c^{XZ}_{\alpha\beta}$ with  ${\alpha \beta}$ = $e\mu$, $e\tau$, and  $\mu\tau$) and mixing angle $\theta_{23}$ at 2$\sigma$, 2.5$\sigma$, and 3$\sigma$ CL.}
    \label{corelation_th}
    \end{figure}
     \end{minipage}
     
    \begin{figure}[H]
    \includegraphics[height= 0.5\textwidth,width=1.0\textwidth]{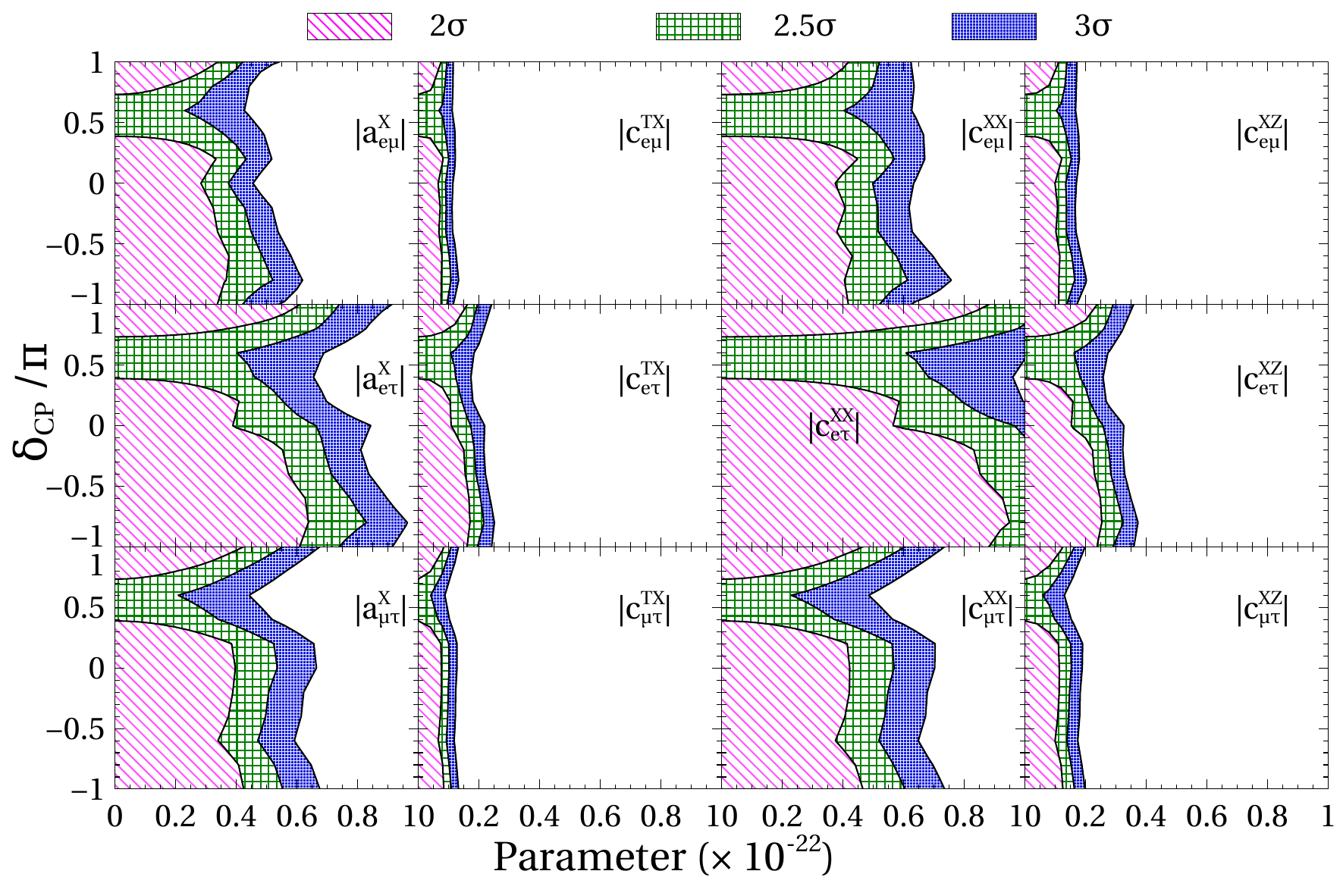}
    \caption{Correlations between the non-diagonal parameters ($a^{X}_{\alpha\beta}$, $c^{TX}_{\alpha\beta}$, $c^{XX}_{\alpha\beta}$, $c^{XZ}_{\alpha\beta}$ with  ${\alpha \beta}$ = $e\mu$, $e\tau$, and  $\mu\tau$)  and Dirac cp-phase $\delta_{CP}$ at 2$\sigma$, 2.5$\sigma$, and 3$\sigma$ CL.}
    \label{corelation_cp}
   \end{figure}

    \begin{figure}[H]
     
    \includegraphics[height= 0.55\textwidth,width=1.0\textwidth]{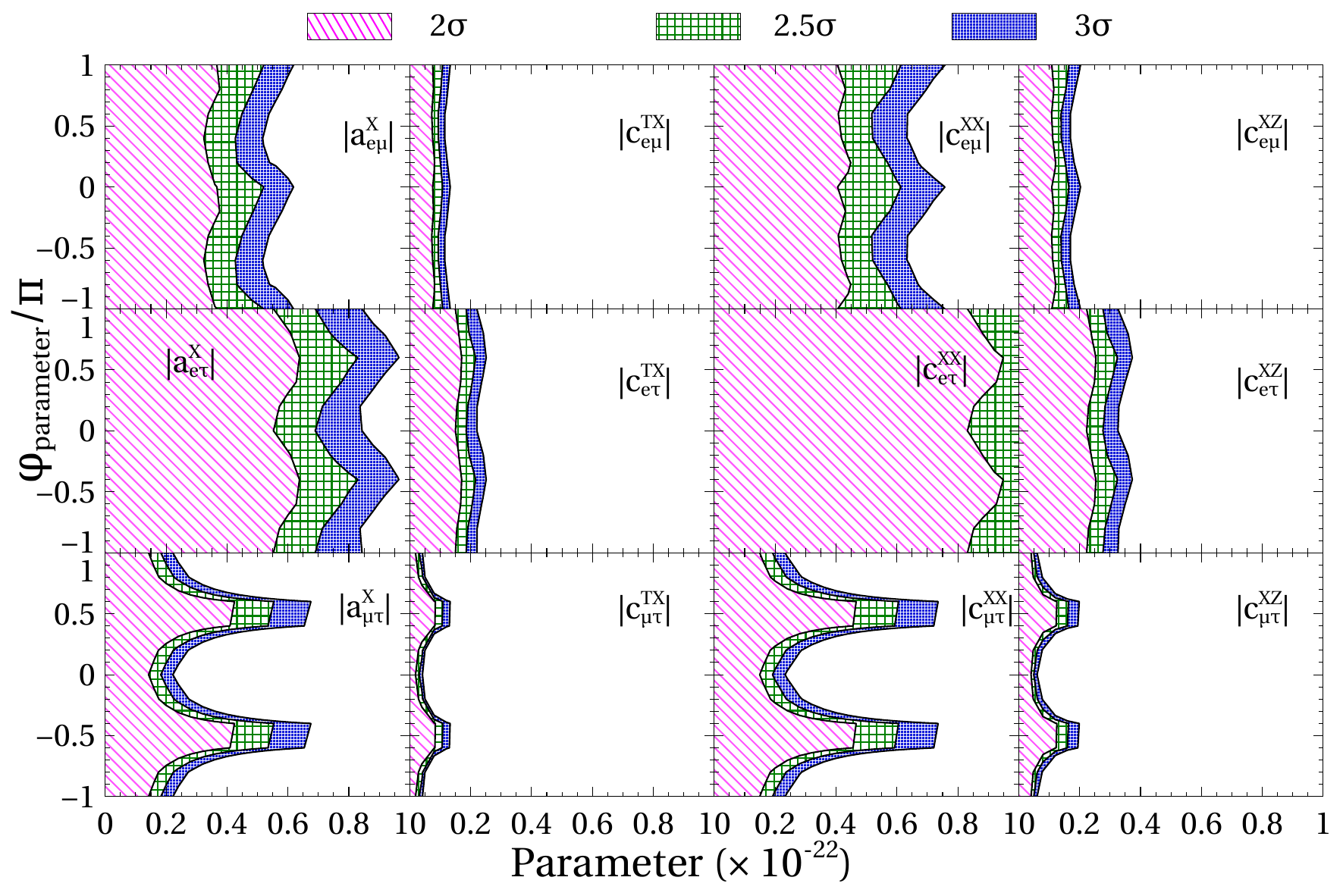}
    \caption{Correlations between the non-diagonal parameters ($a^{X}_{\alpha\beta}$, $c^{TX}_{\alpha\beta}$, $c^{XX}_{\alpha\beta}$, $c^{XZ}_{\alpha\beta}$ with ${\alpha \beta}$ = $e\mu$, $e\tau$, and  $\mu\tau$) at 2$\sigma$, 2.5$\sigma$, and 3$\sigma$ CL}
    \label{corelation_ph}  
   \end{figure}

    \begin{figure}[H]
      \includegraphics[height= 0.60\textwidth,width=1.0\textwidth]{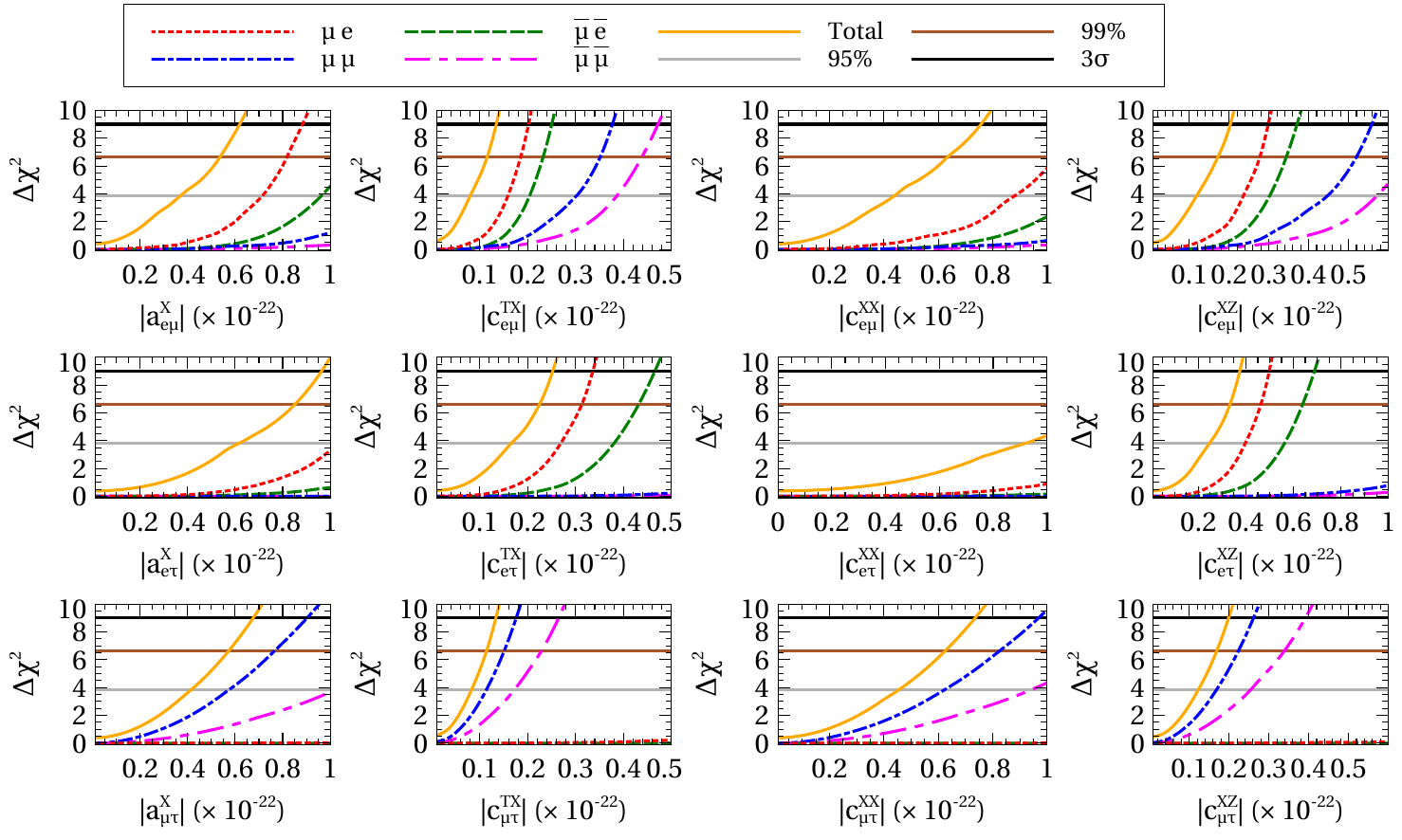}
      \caption{Sensitivity plots of the LIV parameters for $\mu $ $ e$, $\mu $ $\mu$, $\bar{\mu}$ $ \bar{e}$, $\bar{\mu} $ $ \bar{\mu}$ and all channels combined with $2\sigma$ and $3\sigma$ cut}
      \label{sensitivity}
    \end{figure}
\end{widetext}
\begin{widetext}
\begin{center}
\begin{table}[h!]
  \caption{Summary of upper limits at 95\% and  99.7\% C.L . for all 12 LIV parameter under sidereal anaylsis.}
  \label{table4}
  \setlength{\tabcolsep}{21pt}
 \renewcommand{\arraystretch}{1.65}     
  \begin{center}
  \begin{tabular}{|c|c|c|c|c|}
    \hline
        Parameter &  Previous limit & References & This work & This work \\ 
              &      $3\sigma$          &       & 95\% C.L. & 99.7\% C.L. \\ \hline
   $|a^{X}_{e\mu}| = |a^{Y}_{e\mu}| $ & $2.2\times 10^{-20}$  & ~\cite{MINOS:2008fnv}~\cite{MINOS:2012ozn} & $3.68 \times 10^{-23}$  & $6.18 \times 10^{-23}$\\
    $|a^{X}_{e\tau}| = |a^{Y}_{e\tau}|$  & $ NA $ & & $6.2 \times 10^{-23}$  & $9.64 \times 10^{-23}$ \\
    $ |a^{X}_{\mu \tau}| = |a^{Y}_{\mu \tau}| $  & $ 1.8\times 10^{-23} $ & ~\cite{IceCube:2010fyu} & $4.13 \times 10^{-23}$ & $ 6.75 \times 10^{-23} $ \\ \hline
        
    $ |c^{TX}_{e\mu}| = |c^{TY}_{e\mu}| $  & $ 9.0\times 10^{-23} $ & ~\cite{MINOS:2008fnv}~\cite{MINOS:2012ozn} &  $ 8.0\times 10^{-24}  $ & $ 1.32\times 10^{-23}$ \\
    $ |c^{TX}_{e\tau}| = |c^{TY}_{e\tau}| $  & $ NA $ & & $ 1.66\times 10^{-23}  $ & $ 2.5\times 10^{-23} $ \\
    $ |c^{TX}_{\mu \tau}| = |c^{TY}_{\mu \tau}| $  & $ 3.7\times 10^{-27} $ &  ~\cite{IceCube:2010fyu} & $ 8.2\times 10^{-24}  $ & $ 1.32\times 10^{-23}$ \\ \hline
    
    $ |c^{XX}_{e\mu}| = |c^{YY}_{e\mu}|$  & $ 4.6\times 10^{-21} $ & ~\cite{MINOS:2008fnv}~\cite{MINOS:2012ozn} & $ 4.38\times 10^{-23}  $ & $ 7.57\times 10^{-23}$ \\
    $ |c^{XX}_{e\tau}| = |c^{YY}_{e\tau}|$  & $ NA $ & & $ 9.26\times 10^{-23}  $ & $ -- $ \\
    $ |c^{XX}_{\mu \tau}| = |c^{YY}_{\mu \tau}| $  & $ 2.5\times 10^{-23} $ & ~\cite{MINOS:2010kat} & $ 4.54\times 10^{-23}  $ & $ 7.35\times 10^{-23}$ \\ \hline
    
    $ |c^{XZ}_{e\mu}| = |c^{YZ}_{e\mu}|$  & $ 1.1\times 10^{-21} $ & ~\cite{MINOS:2008fnv}~\cite{MINOS:2012ozn} & $ 1.1\times 10^{-23}  $ & $ 2.04\times 10^{-23}$ \\
    $ |c^{XZ}_{e\tau}| = |c^{XZ}_{e\tau}|$  & $ NA $ & & $ 2.46\times 10^{-23}  $ & $ 3.72\times 10^{-23} $ \\
    $ |c^{XZ}_{\mu \tau}| = |c^{XZ}_{\mu \tau}|$  & $ 0.7\times 10^{-23} $ & ~\cite{MINOS:2010kat} & $ 1.21\times 10^{-23}  $ & $ 1.98\times 10^{-23}$ \\ \hline
  \end{tabular}
 \end{center}
\end{table}
\end{center}
\end{widetext}
  We now present the sensitivity of $\nova$-experiment towards constraining the non-diagonal parameters $a^{X}_{\alpha\beta}$,$c^{TX}_{\alpha\beta}$, $c^{XX}_{\alpha\beta}$, $c^{XZ}_{\alpha\beta}$ with  ${\alpha \beta}$ = $e\mu$, $e\tau$, and  $\mu\tau$ using the sidereal analysis. With marginalisation over $\theta_{23}$, $\delta_{CP}$, and $\phi_{\textrm{parameter}}$, the degeneracy effect of these parameter is eliminated. Figure~\ref{sensitivity} illustrates the $\chi^2$ sensitivity of the LIV parameters in both the appearance and disappearance channels, considering both neutrino and anti-neutrino modes.

By adopting a one-parameter-at-a-time analysis, the upper limits at $3\sigma$ level of all 12 LIV parmeters is listed in Table~\ref{table4}.  We note that sidereal analysis with FAR detector provide more stringent constraints on $3\sigma$ level for CPT-violating coefficient $a^{X}_{e\mu}$ ($a^{Y}_{e\mu}$) which are suppressed by 3 orders from previous reported results. In this analysis, we present the first time constraint on non-diagonal LIV parameters LIV parameter corresponding $e\tau$ coefficient which have never been reported previously by any neutrino experiment. Only specific channels have been used in previous studies of the sidereal impact in neutrino sectors. We note that $\nova$ is not able to improve the results on $\mu \tau$ parameters from existing bounds, as the neutrino beam used in $\nova$ is not able to create $\tau$ hence it is not sensitive for $\mu \rightarrow \tau$ channel.


\section{Summary and conclusion}
\label{sec::summary}
The presented work focuses on investigating Lorentz Invariance Violation (LIV) through the sidereal effect within the context of the $\nova$ experiment. Oscillation probabilities and events are simulated using the GLoBES software with desired experimental configurations. This analysis examines the influence of the sidereal effect on various LIV parameters within the oscillation probability spectra.

As the sidereal effect is time-dependent, the flux variation with sidereal time may alter the event to LST  spectra. Since there is no prior experimental data available on flux variation with LST, an average constant flux over the entire sidereal period is considered. It is observed that Eq.~\ref{pmue}~and~\ref{pmumu} accurately describe the sidereal effect upto leading order  for the long-baseline scenario. Furthermore, it is demonstrated that LIV parameters exhibit complementary characteristics in the appearance and disappearance channels. Certain parameters predominantly affect the appearance channel, while others primarily impact the disappearance channel. This pattern is also reflected in the sensitivity analysis, as sensitivity is specific to each channel.

By combining the effects from all channels, the $\nova$ experiment(FD) can provide new constraints on LIV parameter values at a $3\sigma$ confidence level. The study indicates that the $\nova$ experiment(FD) has the capability to detect and constrain sidereal effects effectively using the far detector. However, not all parameters could be explored with improved limits.

Since non-isotropic LIV is direction-dependent, it cannot be constrained through conventional neutrino oscillation studies. Figs ~\ref{corelation_cp} and ~\ref{corelation_th} illustrate that the sidereal effect is sensitive to the standard oscillation parameters $\theta_{23}$ and $\delta_{CP}$. Uncertainties in these parameters can reduce the sensitivity of $\nova$ to sidereal parameters.

Moreover, the sidereal parameters are highly influenced by the baseline length and neutrino energy. Future long-baseline experiments with longer baselines and higher energies, such as DUNE, T2HKK and P2O may offer enhanced sensitivity to non-isotropic LIV parameters. 

\section{Acknowledgments}
We acknowledge financial support from the DST, New Delhi, India for providing funds under the Umbrella Scheme Research and Development (S. Mishra, S. Shukla, and V. Singh), CSIR, New Delhi, India (S. Shukla) and  UGC-BSR Research Start Up Grant, India Contract F.30-584/2021 (BSR) (L. S.). We would like to thank Dr. M. Masud for many insightful discussions. 

\bibliography{LIVNovaRef}

\begin{thebibliography}{41}%
\makeatletter
\providecommand \@ifxundefined [1]{%
 \@ifx{#1\undefined}
}%
\providecommand \@ifnum [1]{%
 \ifnum #1\expandafter \@firstoftwo
 \else \expandafter \@secondoftwo
 \fi
}%
\providecommand \@ifx [1]{%
 \ifx #1\expandafter \@firstoftwo
 \else \expandafter \@secondoftwo
 \fi
}%
\providecommand \natexlab [1]{#1}%
\providecommand \enquote  [1]{``#1''}%
\providecommand \bibnamefont  [1]{#1}%
\providecommand \bibfnamefont [1]{#1}%
\providecommand \citenamefont [1]{#1}%
\providecommand \href@noop [0]{\@secondoftwo}%
\providecommand \href [0]{\begingroup \@sanitize@url \@href}%
\providecommand \@href[1]{\@@startlink{#1}\@@href}%
\providecommand \@@href[1]{\endgroup#1\@@endlink}%
\providecommand \@sanitize@url [0]{\catcode `\\12\catcode `\$12\catcode
  `\&12\catcode `\#12\catcode `\^12\catcode `\_12\catcode `\%12\relax}%
\providecommand \@@startlink[1]{}%
\providecommand \@@endlink[0]{}%
\providecommand \url  [0]{\begingroup\@sanitize@url \@url }%
\providecommand \@url [1]{\endgroup\@href {#1}{\urlprefix }}%
\providecommand \urlprefix  [0]{URL }%
\providecommand \Eprint [0]{\href }%
\providecommand \doibase [0]{https://doi.org/}%
\providecommand \selectlanguage [0]{\@gobble}%
\providecommand \bibinfo  [0]{\@secondoftwo}%
\providecommand \bibfield  [0]{\@secondoftwo}%
\providecommand \translation [1]{[#1]}%
\providecommand \BibitemOpen [0]{}%
\providecommand \bibitemStop [0]{}%
\providecommand \bibitemNoStop [0]{.\EOS\space}%
\providecommand \EOS [0]{\spacefactor3000\relax}%
\providecommand \BibitemShut  [1]{\csname bibitem#1\endcsname}%
\let\auto@bib@innerbib\@empty
\bibitem [{\citenamefont {Kostelecky}\ and\ \citenamefont
  {Samuel}(1989)}]{Kostelecky:1988zi}%
  \BibitemOpen
  \bibfield  {author} {\bibinfo {author} {\bibfnamefont {V.~A.}\ \bibnamefont
  {Kostelecky}}\ and\ \bibinfo {author} {\bibfnamefont {S.}~\bibnamefont
  {Samuel}},\ }\bibfield  {title} {\bibinfo {title} {{Spontaneous Breaking of
  Lorentz Symmetry in String Theory}},\ }\href
  {https://doi.org/10.1103/PhysRevD.39.683} {\bibfield  {journal} {\bibinfo
  {journal} {Phys. Rev. D}\ }\textbf {\bibinfo {volume} {39}},\ \bibinfo
  {pages} {683} (\bibinfo {year} {1989})}\BibitemShut {NoStop}%
\bibitem [{\citenamefont {Kostelecky}\ and\ \citenamefont
  {Potting}(1991)}]{Kostelecky:1991ak}%
  \BibitemOpen
  \bibfield  {author} {\bibinfo {author} {\bibfnamefont {V.~A.}\ \bibnamefont
  {Kostelecky}}\ and\ \bibinfo {author} {\bibfnamefont {R.}~\bibnamefont
  {Potting}},\ }\bibfield  {title} {\bibinfo {title} {{CPT and strings}},\
  }\href {https://doi.org/10.1016/0550-3213(91)90071-5} {\bibfield  {journal}
  {\bibinfo  {journal} {Nucl. Phys. B}\ }\textbf {\bibinfo {volume} {359}},\
  \bibinfo {pages} {545} (\bibinfo {year} {1991})}\BibitemShut {NoStop}%
\bibitem [{\citenamefont {Gambini}\ and\ \citenamefont
  {Pullin}(1999)}]{Gambini:1998it}%
  \BibitemOpen
  \bibfield  {author} {\bibinfo {author} {\bibfnamefont {R.}~\bibnamefont
  {Gambini}}\ and\ \bibinfo {author} {\bibfnamefont {J.}~\bibnamefont
  {Pullin}},\ }\bibfield  {title} {\bibinfo {title} {{Nonstandard optics from
  quantum space-time}},\ }\href {https://doi.org/10.1103/PhysRevD.59.124021}
  {\bibfield  {journal} {\bibinfo  {journal} {Phys. Rev. D}\ }\textbf {\bibinfo
  {volume} {59}},\ \bibinfo {pages} {124021} (\bibinfo {year} {1999})},\
  \Eprint {https://arxiv.org/abs/gr-qc/9809038} {arXiv:gr-qc/9809038}
  \BibitemShut {NoStop}%
\bibitem [{\citenamefont {Burgess}\ \emph {et~al.}(2002)\citenamefont
  {Burgess}, \citenamefont {Cline}, \citenamefont {Filotas}, \citenamefont
  {Matias},\ and\ \citenamefont {Moore}}]{CliffP.Burgess_2002}%
  \BibitemOpen
  \bibfield  {author} {\bibinfo {author} {\bibfnamefont {C.~P.}\ \bibnamefont
  {Burgess}}, \bibinfo {author} {\bibfnamefont {J.~M.}\ \bibnamefont {Cline}},
  \bibinfo {author} {\bibfnamefont {E.}~\bibnamefont {Filotas}}, \bibinfo
  {author} {\bibfnamefont {J.}~\bibnamefont {Matias}},\ and\ \bibinfo {author}
  {\bibfnamefont {G.~D.}\ \bibnamefont {Moore}},\ }\bibfield  {title} {\bibinfo
  {title} {Loop-generated bounds on changes to the graviton dispersion
  relation},\ }\href {https://doi.org/10.1088/1126-6708/2002/03/043} {\bibfield
   {journal} {\bibinfo  {journal} {Journal of High Energy Physics}\ }\textbf
  {\bibinfo {volume} {2002}},\ \bibinfo {pages} {043} (\bibinfo {year}
  {2002})}\BibitemShut {NoStop}%
\bibitem [{\citenamefont {Sahoo}\ \emph {et~al.}(2022)\citenamefont {Sahoo},
  \citenamefont {Kumar},\ and\ \citenamefont {Agarwalla}}]{Sahoo:2021dit}%
  \BibitemOpen
  \bibfield  {author} {\bibinfo {author} {\bibfnamefont {S.}~\bibnamefont
  {Sahoo}}, \bibinfo {author} {\bibfnamefont {A.}~\bibnamefont {Kumar}},\ and\
  \bibinfo {author} {\bibfnamefont {S.~K.}\ \bibnamefont {Agarwalla}},\
  }\bibfield  {title} {\bibinfo {title} {{Probing Lorentz Invariance Violation
  with atmospheric neutrinos at INO-ICAL}},\ }\href
  {https://doi.org/10.1007/JHEP03(2022)050} {\bibfield  {journal} {\bibinfo
  {journal} {JHEP}\ }\textbf {\bibinfo {volume} {03}},\ \bibinfo {pages}
  {050}},\ \Eprint {https://arxiv.org/abs/2110.13207} {arXiv:2110.13207
  [hep-ph]} \BibitemShut {NoStop}%
\bibitem [{\citenamefont {Colladay}\ and\ \citenamefont
  {Kostelecky}(1998)}]{Colladay:1998fq}%
  \BibitemOpen
  \bibfield  {author} {\bibinfo {author} {\bibfnamefont {D.}~\bibnamefont
  {Colladay}}\ and\ \bibinfo {author} {\bibfnamefont {V.~A.}\ \bibnamefont
  {Kostelecky}},\ }\bibfield  {title} {\bibinfo {title} {{Lorentz violating
  extension of the standard model}},\ }\href
  {https://doi.org/10.1103/PhysRevD.58.116002} {\bibfield  {journal} {\bibinfo
  {journal} {Phys. Rev. D}\ }\textbf {\bibinfo {volume} {58}},\ \bibinfo
  {pages} {116002} (\bibinfo {year} {1998})},\ \Eprint
  {https://arxiv.org/abs/hep-ph/9809521} {arXiv:hep-ph/9809521} \BibitemShut
  {NoStop}%
\bibitem [{\citenamefont {Bluhm}(2006)}]{Bluhm:2005uj}%
  \BibitemOpen
  \bibfield  {author} {\bibinfo {author} {\bibfnamefont {R.}~\bibnamefont
  {Bluhm}},\ }\bibfield  {title} {\bibinfo {title} {{Overview of the SME:
  Implications and phenomenology of Lorentz violation}},\ }\href
  {https://doi.org/10.1007/3-540-34523-X_8} {\bibfield  {journal} {\bibinfo
  {journal} {Lect. Notes Phys.}\ }\textbf {\bibinfo {volume} {702}},\ \bibinfo
  {pages} {191} (\bibinfo {year} {2006})},\ \Eprint
  {https://arxiv.org/abs/hep-ph/0506054} {arXiv:hep-ph/0506054} \BibitemShut
  {NoStop}%
\bibitem [{\citenamefont {Kosteleck\'y}\ and\ \citenamefont
  {Russell}(2011)}]{RevModPhys.83.11}%
  \BibitemOpen
  \bibfield  {author} {\bibinfo {author} {\bibfnamefont {V.~A.}\ \bibnamefont
  {Kosteleck\'y}}\ and\ \bibinfo {author} {\bibfnamefont {N.}~\bibnamefont
  {Russell}},\ }\bibfield  {title} {\bibinfo {title} {Data tables for lorentz
  and $cpt$ violation},\ }\href {https://doi.org/10.1103/RevModPhys.83.11}
  {\bibfield  {journal} {\bibinfo  {journal} {Rev. Mod. Phys.}\ }\textbf
  {\bibinfo {volume} {83}},\ \bibinfo {pages} {11} (\bibinfo {year}
  {2011})}\BibitemShut {NoStop}%
\bibitem [{\citenamefont {Fukuda}\ \emph {et~al.}(1998)\citenamefont {Fukuda}
  \emph {et~al.}}]{Super-Kamiokande:1998qwk}%
  \BibitemOpen
  \bibfield  {author} {\bibinfo {author} {\bibfnamefont {Y.}~\bibnamefont
  {Fukuda}} \emph {et~al.} (\bibinfo {collaboration} {Super-Kamiokande}),\
  }\bibfield  {title} {\bibinfo {title} {{Measurements of the solar neutrino
  flux from Super-Kamiokande's first 300 days}},\ }\href
  {https://doi.org/10.1103/PhysRevLett.81.1158} {\bibfield  {journal} {\bibinfo
   {journal} {Phys. Rev. Lett.}\ }\textbf {\bibinfo {volume} {81}},\ \bibinfo
  {pages} {1158} (\bibinfo {year} {1998})},\ \bibinfo {note} {[Erratum:
  Phys.Rev.Lett. 81, 4279 (1998)]},\ \Eprint
  {https://arxiv.org/abs/hep-ex/9805021} {arXiv:hep-ex/9805021} \BibitemShut
  {NoStop}%
\bibitem [{\citenamefont {Eguchi}\ \emph {et~al.}(2003)\citenamefont {Eguchi}
  \emph {et~al.}}]{KamLAND:2002uet}%
  \BibitemOpen
  \bibfield  {author} {\bibinfo {author} {\bibfnamefont {K.}~\bibnamefont
  {Eguchi}} \emph {et~al.} (\bibinfo {collaboration} {KamLAND}),\ }\bibfield
  {title} {\bibinfo {title} {{First results from KamLAND: Evidence for reactor
  anti-neutrino disappearance}},\ }\href
  {https://doi.org/10.1103/PhysRevLett.90.021802} {\bibfield  {journal}
  {\bibinfo  {journal} {Phys. Rev. Lett.}\ }\textbf {\bibinfo {volume} {90}},\
  \bibinfo {pages} {021802} (\bibinfo {year} {2003})},\ \Eprint
  {https://arxiv.org/abs/hep-ex/0212021} {arXiv:hep-ex/0212021} \BibitemShut
  {NoStop}%
\bibitem [{\citenamefont {Ahn}\ \emph {et~al.}(2006)\citenamefont {Ahn} \emph
  {et~al.}}]{K2K:2006yov}%
  \BibitemOpen
  \bibfield  {author} {\bibinfo {author} {\bibfnamefont {M.~H.}\ \bibnamefont
  {Ahn}} \emph {et~al.} (\bibinfo {collaboration} {K2K}),\ }\bibfield  {title}
  {\bibinfo {title} {{Measurement of Neutrino Oscillation by the K2K
  Experiment}},\ }\href {https://doi.org/10.1103/PhysRevD.74.072003} {\bibfield
   {journal} {\bibinfo  {journal} {Phys. Rev. D}\ }\textbf {\bibinfo {volume}
  {74}},\ \bibinfo {pages} {072003} (\bibinfo {year} {2006})},\ \Eprint
  {https://arxiv.org/abs/hep-ex/0606032} {arXiv:hep-ex/0606032} \BibitemShut
  {NoStop}%
\bibitem [{\citenamefont {Adamson}\ \emph {et~al.}(2008)\citenamefont {Adamson}
  \emph {et~al.}}]{MINOS:2008fnv}%
  \BibitemOpen
  \bibfield  {author} {\bibinfo {author} {\bibfnamefont {P.}~\bibnamefont
  {Adamson}} \emph {et~al.} (\bibinfo {collaboration} {MINOS}),\ }\bibfield
  {title} {\bibinfo {title} {{Testing Lorentz Invariance and CPT Conservation
  with NuMI Neutrinos in the MINOS Near Detector}},\ }\href
  {https://doi.org/10.1103/PhysRevLett.101.151601} {\bibfield  {journal}
  {\bibinfo  {journal} {Phys. Rev. Lett.}\ }\textbf {\bibinfo {volume} {101}},\
  \bibinfo {pages} {151601} (\bibinfo {year} {2008})},\ \Eprint
  {https://arxiv.org/abs/0806.4945} {arXiv:0806.4945 [hep-ex]} \BibitemShut
  {NoStop}%
\bibitem [{\citenamefont {Adamson}\ \emph {et~al.}(2010)\citenamefont {Adamson}
  \emph {et~al.}}]{MINOS:2010kat}%
  \BibitemOpen
  \bibfield  {author} {\bibinfo {author} {\bibfnamefont {P.}~\bibnamefont
  {Adamson}} \emph {et~al.} (\bibinfo {collaboration} {MINOS}),\ }\bibfield
  {title} {\bibinfo {title} {{A Search for Lorentz Invariance and CPT Violation
  with the MINOS Far Detector}},\ }\href
  {https://doi.org/10.1103/PhysRevLett.105.151601} {\bibfield  {journal}
  {\bibinfo  {journal} {Phys. Rev. Lett.}\ }\textbf {\bibinfo {volume} {105}},\
  \bibinfo {pages} {151601} (\bibinfo {year} {2010})},\ \Eprint
  {https://arxiv.org/abs/1007.2791} {arXiv:1007.2791 [hep-ex]} \BibitemShut
  {NoStop}%
\bibitem [{\citenamefont {Adamson}\ \emph {et~al.}(2012)\citenamefont {Adamson}
  \emph {et~al.}}]{MINOS:2012ozn}%
  \BibitemOpen
  \bibfield  {author} {\bibinfo {author} {\bibfnamefont {P.}~\bibnamefont
  {Adamson}} \emph {et~al.} (\bibinfo {collaboration} {MINOS}),\ }\bibfield
  {title} {\bibinfo {title} {{Search for Lorentz invariance and CPT violation
  with muon antineutrinos in the MINOS Near Detector}},\ }\href
  {https://doi.org/10.1103/PhysRevD.85.031101} {\bibfield  {journal} {\bibinfo
  {journal} {Phys. Rev. D}\ }\textbf {\bibinfo {volume} {85}},\ \bibinfo
  {pages} {031101} (\bibinfo {year} {2012})},\ \Eprint
  {https://arxiv.org/abs/1201.2631} {arXiv:1201.2631 [hep-ex]} \BibitemShut
  {NoStop}%
\bibitem [{\citenamefont {Abbasi}\ \emph {et~al.}(2010)\citenamefont {Abbasi}
  \emph {et~al.}}]{IceCube:2010fyu}%
  \BibitemOpen
  \bibfield  {author} {\bibinfo {author} {\bibfnamefont {R.}~\bibnamefont
  {Abbasi}} \emph {et~al.} (\bibinfo {collaboration} {IceCube}),\ }\bibfield
  {title} {\bibinfo {title} {{Search for a Lorentz-violating sidereal signal
  with atmospheric neutrinos in IceCube}},\ }\href
  {https://doi.org/10.1103/PhysRevD.82.112003} {\bibfield  {journal} {\bibinfo
  {journal} {Phys. Rev. D}\ }\textbf {\bibinfo {volume} {82}},\ \bibinfo
  {pages} {112003} (\bibinfo {year} {2010})},\ \Eprint
  {https://arxiv.org/abs/1010.4096} {arXiv:1010.4096 [astro-ph.HE]}
  \BibitemShut {NoStop}%
\bibitem [{\citenamefont {Auerbach}\ \emph {et~al.}(2005)\citenamefont
  {Auerbach} \emph {et~al.}}]{LSND:2005oop}%
  \BibitemOpen
  \bibfield  {author} {\bibinfo {author} {\bibfnamefont {L.~B.}\ \bibnamefont
  {Auerbach}} \emph {et~al.} (\bibinfo {collaboration} {LSND}),\ }\bibfield
  {title} {\bibinfo {title} {{Tests of Lorentz violation in anti-nu(mu)
  ---\ensuremath{>} anti-nu(e) oscillations}},\ }\href
  {https://doi.org/10.1103/PhysRevD.72.076004} {\bibfield  {journal} {\bibinfo
  {journal} {Phys. Rev. D}\ }\textbf {\bibinfo {volume} {72}},\ \bibinfo
  {pages} {076004} (\bibinfo {year} {2005})},\ \Eprint
  {https://arxiv.org/abs/hep-ex/0506067} {arXiv:hep-ex/0506067} \BibitemShut
  {NoStop}%
\bibitem [{\citenamefont {Abe}\ \emph {et~al.}(2015)\citenamefont {Abe} \emph
  {et~al.}}]{Super-Kamiokande:2014exs}%
  \BibitemOpen
  \bibfield  {author} {\bibinfo {author} {\bibfnamefont {K.}~\bibnamefont
  {Abe}} \emph {et~al.} (\bibinfo {collaboration} {Super-Kamiokande}),\
  }\bibfield  {title} {\bibinfo {title} {{Test of Lorentz invariance with
  atmospheric neutrinos}},\ }\href {https://doi.org/10.1103/PhysRevD.91.052003}
  {\bibfield  {journal} {\bibinfo  {journal} {Phys. Rev. D}\ }\textbf {\bibinfo
  {volume} {91}},\ \bibinfo {pages} {052003} (\bibinfo {year} {2015})},\
  \Eprint {https://arxiv.org/abs/1410.4267} {arXiv:1410.4267 [hep-ex]}
  \BibitemShut {NoStop}%
\bibitem [{\citenamefont {Abe}\ \emph {et~al.}(2017)\citenamefont {Abe} \emph
  {et~al.}}]{T2K:2017ega}%
  \BibitemOpen
  \bibfield  {author} {\bibinfo {author} {\bibfnamefont {K.}~\bibnamefont
  {Abe}} \emph {et~al.} (\bibinfo {collaboration} {T2K}),\ }\bibfield  {title}
  {\bibinfo {title} {{Search for Lorentz and CPT violation using sidereal time
  dependence of neutrino flavor transitions over a short baseline}},\ }\href
  {https://doi.org/10.1103/PhysRevD.95.111101} {\bibfield  {journal} {\bibinfo
  {journal} {Phys. Rev. D}\ }\textbf {\bibinfo {volume} {95}},\ \bibinfo
  {pages} {111101} (\bibinfo {year} {2017})},\ \Eprint
  {https://arxiv.org/abs/1703.01361} {arXiv:1703.01361 [hep-ex]} \BibitemShut
  {NoStop}%
\bibitem [{\citenamefont {Adey}\ \emph {et~al.}(2018)\citenamefont {Adey} \emph
  {et~al.}}]{DayaBay:2018fsh}%
  \BibitemOpen
  \bibfield  {author} {\bibinfo {author} {\bibfnamefont {D.}~\bibnamefont
  {Adey}} \emph {et~al.} (\bibinfo {collaboration} {Daya Bay}),\ }\bibfield
  {title} {\bibinfo {title} {{Search for a time-varying electron antineutrino
  signal at Daya Bay}},\ }\href {https://doi.org/10.1103/PhysRevD.98.092013}
  {\bibfield  {journal} {\bibinfo  {journal} {Phys. Rev. D}\ }\textbf {\bibinfo
  {volume} {98}},\ \bibinfo {pages} {092013} (\bibinfo {year} {2018})},\
  \Eprint {https://arxiv.org/abs/1809.04660} {arXiv:1809.04660 [hep-ex]}
  \BibitemShut {NoStop}%
\bibitem [{\citenamefont {Aguilar-Arevalo}\ \emph {et~al.}(2013)\citenamefont
  {Aguilar-Arevalo} \emph {et~al.}}]{MiniBooNE:2011pix}%
  \BibitemOpen
  \bibfield  {author} {\bibinfo {author} {\bibfnamefont {A.~A.}\ \bibnamefont
  {Aguilar-Arevalo}} \emph {et~al.} (\bibinfo {collaboration} {MiniBooNE}),\
  }\bibfield  {title} {\bibinfo {title} {{Test of Lorentz and CPT violation
  with Short Baseline Neutrino Oscillation Excesses}},\ }\href
  {https://doi.org/10.1016/j.physletb.2012.12.020} {\bibfield  {journal}
  {\bibinfo  {journal} {Phys. Lett. B}\ }\textbf {\bibinfo {volume} {718}},\
  \bibinfo {pages} {1303} (\bibinfo {year} {2013})},\ \Eprint
  {https://arxiv.org/abs/1109.3480} {arXiv:1109.3480 [hep-ex]} \BibitemShut
  {NoStop}%
\bibitem [{\citenamefont {Abe}\ \emph {et~al.}(2012)\citenamefont {Abe} \emph
  {et~al.}}]{DoubleChooz:2012eiq}%
  \BibitemOpen
  \bibfield  {author} {\bibinfo {author} {\bibfnamefont {Y.}~\bibnamefont
  {Abe}} \emph {et~al.} (\bibinfo {collaboration} {Double Chooz}),\ }\bibfield
  {title} {\bibinfo {title} {{First Test of Lorentz Violation with a
  Reactor-based Antineutrino Experiment}},\ }\href
  {https://doi.org/10.1103/PhysRevD.86.112009} {\bibfield  {journal} {\bibinfo
  {journal} {Phys. Rev. D}\ }\textbf {\bibinfo {volume} {86}},\ \bibinfo
  {pages} {112009} (\bibinfo {year} {2012})},\ \Eprint
  {https://arxiv.org/abs/1209.5810} {arXiv:1209.5810 [hep-ex]} \BibitemShut
  {NoStop}%
\bibitem [{\citenamefont {Barenboim}\ \emph {et~al.}(2019)\citenamefont
  {Barenboim}, \citenamefont {Masud}, \citenamefont {Ternes},\ and\
  \citenamefont {T\'ortola}}]{Barenboim:2018ctx}%
  \BibitemOpen
  \bibfield  {author} {\bibinfo {author} {\bibfnamefont {G.}~\bibnamefont
  {Barenboim}}, \bibinfo {author} {\bibfnamefont {M.}~\bibnamefont {Masud}},
  \bibinfo {author} {\bibfnamefont {C.~A.}\ \bibnamefont {Ternes}},\ and\
  \bibinfo {author} {\bibfnamefont {M.}~\bibnamefont {T\'ortola}},\ }\bibfield
  {title} {\bibinfo {title} {{Exploring the intrinsic Lorentz-violating
  parameters at DUNE}},\ }\href
  {https://doi.org/10.1016/j.physletb.2018.11.040} {\bibfield  {journal}
  {\bibinfo  {journal} {Phys. Lett. B}\ }\textbf {\bibinfo {volume} {788}},\
  \bibinfo {pages} {308} (\bibinfo {year} {2019})},\ \Eprint
  {https://arxiv.org/abs/1805.11094} {arXiv:1805.11094 [hep-ph]} \BibitemShut
  {NoStop}%
\bibitem [{\citenamefont {Acero}\ \emph {et~al.}(2019)\citenamefont {Acero}
  \emph {et~al.}}]{NOvA:2019cyt}%
  \BibitemOpen
  \bibfield  {author} {\bibinfo {author} {\bibfnamefont {M.~A.}\ \bibnamefont
  {Acero}} \emph {et~al.} (\bibinfo {collaboration} {NOvA}),\ }\bibfield
  {title} {\bibinfo {title} {{First Measurement of Neutrino Oscillation
  Parameters using Neutrinos and Antineutrinos by NOvA}},\ }\href
  {https://doi.org/10.1103/PhysRevLett.123.151803} {\bibfield  {journal}
  {\bibinfo  {journal} {Phys. Rev. Lett.}\ }\textbf {\bibinfo {volume} {123}},\
  \bibinfo {pages} {151803} (\bibinfo {year} {2019})},\ \Eprint
  {https://arxiv.org/abs/1906.04907} {arXiv:1906.04907 [hep-ex]} \BibitemShut
  {NoStop}%
\bibitem [{\citenamefont {Diaz}\ \emph {et~al.}(2009)\citenamefont {Diaz},
  \citenamefont {Kostelecky},\ and\ \citenamefont {Mewes}}]{Diaz:2009qk}%
  \BibitemOpen
  \bibfield  {author} {\bibinfo {author} {\bibfnamefont {J.~S.}\ \bibnamefont
  {Diaz}}, \bibinfo {author} {\bibfnamefont {V.~A.}\ \bibnamefont
  {Kostelecky}},\ and\ \bibinfo {author} {\bibfnamefont {M.}~\bibnamefont
  {Mewes}},\ }\bibfield  {title} {\bibinfo {title} {{Perturbative Lorentz and
  CPT violation for neutrino and antineutrino oscillations}},\ }\href
  {https://doi.org/10.1103/PhysRevD.80.076007} {\bibfield  {journal} {\bibinfo
  {journal} {Phys. Rev. D}\ }\textbf {\bibinfo {volume} {80}},\ \bibinfo
  {pages} {076007} (\bibinfo {year} {2009})},\ \Eprint
  {https://arxiv.org/abs/0908.1401} {arXiv:0908.1401 [hep-ph]} \BibitemShut
  {NoStop}%
\bibitem [{\citenamefont {Kopp}\ \emph {et~al.}(2008)\citenamefont {Kopp},
  \citenamefont {Lindner}, \citenamefont {Ota},\ and\ \citenamefont
  {Sato}}]{Kopp:2007ne}%
  \BibitemOpen
  \bibfield  {author} {\bibinfo {author} {\bibfnamefont {J.}~\bibnamefont
  {Kopp}}, \bibinfo {author} {\bibfnamefont {M.}~\bibnamefont {Lindner}},
  \bibinfo {author} {\bibfnamefont {T.}~\bibnamefont {Ota}},\ and\ \bibinfo
  {author} {\bibfnamefont {J.}~\bibnamefont {Sato}},\ }\bibfield  {title}
  {\bibinfo {title} {{Non-standard neutrino interactions in reactor and
  superbeam experiments}},\ }\href {https://doi.org/10.1103/PhysRevD.77.013007}
  {\bibfield  {journal} {\bibinfo  {journal} {Phys. Rev. D}\ }\textbf {\bibinfo
  {volume} {77}},\ \bibinfo {pages} {013007} (\bibinfo {year} {2008})},\
  \Eprint {https://arxiv.org/abs/0708.0152} {arXiv:0708.0152 [hep-ph]}
  \BibitemShut {NoStop}%
\bibitem [{\citenamefont {Kostelecky}\ and\ \citenamefont
  {Mewes}(2004{\natexlab{a}})}]{Kostelecky:2003xn}%
  \BibitemOpen
  \bibfield  {author} {\bibinfo {author} {\bibfnamefont {V.~A.}\ \bibnamefont
  {Kostelecky}}\ and\ \bibinfo {author} {\bibfnamefont {M.}~\bibnamefont
  {Mewes}},\ }\bibfield  {title} {\bibinfo {title} {{Lorentz and CPT violation
  in the neutrino sector}},\ }\href
  {https://doi.org/10.1103/PhysRevD.70.031902} {\bibfield  {journal} {\bibinfo
  {journal} {Phys. Rev. D}\ }\textbf {\bibinfo {volume} {70}},\ \bibinfo
  {pages} {031902} (\bibinfo {year} {2004}{\natexlab{a}})},\ \Eprint
  {https://arxiv.org/abs/hep-ph/0308300} {arXiv:hep-ph/0308300} \BibitemShut
  {NoStop}%
\bibitem [{\citenamefont {Kostelecky}\ and\ \citenamefont
  {Mewes}(2004{\natexlab{b}})}]{Kostelecky:2004hg}%
  \BibitemOpen
  \bibfield  {author} {\bibinfo {author} {\bibfnamefont {V.~A.}\ \bibnamefont
  {Kostelecky}}\ and\ \bibinfo {author} {\bibfnamefont {M.}~\bibnamefont
  {Mewes}},\ }\bibfield  {title} {\bibinfo {title} {{Lorentz violation and
  short-baseline neutrino experiments}},\ }\href
  {https://doi.org/10.1103/PhysRevD.70.076002} {\bibfield  {journal} {\bibinfo
  {journal} {Phys. Rev. D}\ }\textbf {\bibinfo {volume} {70}},\ \bibinfo
  {pages} {076002} (\bibinfo {year} {2004}{\natexlab{b}})},\ \Eprint
  {https://arxiv.org/abs/hep-ph/0406255} {arXiv:hep-ph/0406255} \BibitemShut
  {NoStop}%
\bibitem [{\citenamefont {Liao}\ \emph {et~al.}(2016)\citenamefont {Liao},
  \citenamefont {Marfatia},\ and\ \citenamefont {Whisnant}}]{Liao:2016hsa}%
  \BibitemOpen
  \bibfield  {author} {\bibinfo {author} {\bibfnamefont {J.}~\bibnamefont
  {Liao}}, \bibinfo {author} {\bibfnamefont {D.}~\bibnamefont {Marfatia}},\
  and\ \bibinfo {author} {\bibfnamefont {K.}~\bibnamefont {Whisnant}},\
  }\bibfield  {title} {\bibinfo {title} {{Degeneracies in long-baseline
  neutrino experiments from nonstandard interactions}},\ }\href
  {https://doi.org/10.1103/PhysRevD.93.093016} {\bibfield  {journal} {\bibinfo
  {journal} {Phys. Rev.}\ }\textbf {\bibinfo {volume} {D93}},\ \bibinfo {pages}
  {093016} (\bibinfo {year} {2016})},\ \Eprint
  {https://arxiv.org/abs/1601.00927} {arXiv:1601.00927 [hep-ph]} \BibitemShut
  {NoStop}%
\bibitem [{\citenamefont {Esteves~Chaves}\ \emph {et~al.}(2018)\citenamefont
  {Esteves~Chaves}, \citenamefont {Rossi~Gratieri},\ and\ \citenamefont
  {Peres}}]{Chaves:2018sih}%
  \BibitemOpen
  \bibfield  {author} {\bibinfo {author} {\bibfnamefont {M.}~\bibnamefont
  {Esteves~Chaves}}, \bibinfo {author} {\bibfnamefont {D.}~\bibnamefont
  {Rossi~Gratieri}},\ and\ \bibinfo {author} {\bibfnamefont {O.~L.~G.}\
  \bibnamefont {Peres}},\ }\bibfield  {title} {\bibinfo {title} {{Improvements
  on perturbative oscillation formulas including non-standard neutrino
  Interactions}},\ }\href@noop {} {\  (\bibinfo {year} {2018})},\ \Eprint
  {https://arxiv.org/abs/1810.04979} {arXiv:1810.04979 [hep-ph]} \BibitemShut
  {NoStop}%
\bibitem [{\citenamefont {Dey}\ \emph {et~al.}(2018)\citenamefont {Dey},
  \citenamefont {Nath},\ and\ \citenamefont {Sadhukhan}}]{Dey:2018yht}%
  \BibitemOpen
  \bibfield  {author} {\bibinfo {author} {\bibfnamefont {U.~K.}\ \bibnamefont
  {Dey}}, \bibinfo {author} {\bibfnamefont {N.}~\bibnamefont {Nath}},\ and\
  \bibinfo {author} {\bibfnamefont {S.}~\bibnamefont {Sadhukhan}},\ }\bibfield
  {title} {\bibinfo {title} {{Non-Standard Neutrino Interactions in a Modified
  $\nu$2HDM}},\ }\href {https://doi.org/10.1103/PhysRevD.98.055004} {\bibfield
  {journal} {\bibinfo  {journal} {Phys. Rev.}\ }\textbf {\bibinfo {volume}
  {D98}},\ \bibinfo {pages} {055004} (\bibinfo {year} {2018})},\ \Eprint
  {https://arxiv.org/abs/1804.05808} {arXiv:1804.05808 [hep-ph]} \BibitemShut
  {NoStop}%
\bibitem [{\citenamefont {Yasuda}(2007)}]{Yasuda:2007jp}%
  \BibitemOpen
  \bibfield  {author} {\bibinfo {author} {\bibfnamefont {O.}~\bibnamefont
  {Yasuda}},\ }\bibfield  {title} {\bibinfo {title} {{On the exact formula for
  neutrino oscillation probability by Kimura, Takamura and Yokomakura}},\
  }\href@noop {} {\  (\bibinfo {year} {2007})},\ \Eprint
  {https://arxiv.org/abs/0704.1531} {arXiv:0704.1531 [hep-ph]} \BibitemShut
  {NoStop}%
\bibitem [{\citenamefont {Masud}\ \emph {et~al.}(2016)\citenamefont {Masud},
  \citenamefont {Chatterjee},\ and\ \citenamefont {Mehta}}]{Masud:2015xva}%
  \BibitemOpen
  \bibfield  {author} {\bibinfo {author} {\bibfnamefont {M.}~\bibnamefont
  {Masud}}, \bibinfo {author} {\bibfnamefont {A.}~\bibnamefont {Chatterjee}},\
  and\ \bibinfo {author} {\bibfnamefont {P.}~\bibnamefont {Mehta}},\ }\bibfield
   {title} {\bibinfo {title} {{Probing CP violation signal at DUNE in presence
  of non-standard neutrino interactions}},\ }\href
  {https://doi.org/10.1088/0954-3899/43/9/095005/meta,
  10.1088/0954-3899/43/9/095005} {\bibfield  {journal} {\bibinfo  {journal} {J.
  Phys.}\ }\textbf {\bibinfo {volume} {G43}},\ \bibinfo {pages} {095005}
  (\bibinfo {year} {2016})},\ \Eprint {https://arxiv.org/abs/1510.08261}
  {arXiv:1510.08261 [hep-ph]} \BibitemShut {NoStop}%
\bibitem [{\citenamefont {Masud}\ and\ \citenamefont
  {Mehta}(2016{\natexlab{a}})}]{Masud:2016bvp}%
  \BibitemOpen
  \bibfield  {author} {\bibinfo {author} {\bibfnamefont {M.}~\bibnamefont
  {Masud}}\ and\ \bibinfo {author} {\bibfnamefont {P.}~\bibnamefont {Mehta}},\
  }\bibfield  {title} {\bibinfo {title} {{Nonstandard interactions spoiling the
  CP violation sensitivity at DUNE and other long baseline experiments}},\
  }\href {https://doi.org/10.1103/PhysRevD.94.013014} {\bibfield  {journal}
  {\bibinfo  {journal} {Phys. Rev.}\ }\textbf {\bibinfo {volume} {D94}},\
  \bibinfo {pages} {013014} (\bibinfo {year} {2016}{\natexlab{a}})},\ \Eprint
  {https://arxiv.org/abs/1603.01380} {arXiv:1603.01380 [hep-ph]} \BibitemShut
  {NoStop}%
\bibitem [{\citenamefont {Masud}\ and\ \citenamefont
  {Mehta}(2016{\natexlab{b}})}]{Masud:2016gcl}%
  \BibitemOpen
  \bibfield  {author} {\bibinfo {author} {\bibfnamefont {M.}~\bibnamefont
  {Masud}}\ and\ \bibinfo {author} {\bibfnamefont {P.}~\bibnamefont {Mehta}},\
  }\bibfield  {title} {\bibinfo {title} {{Nonstandard interactions and
  resolving the ordering of neutrino masses at DUNE and other long baseline
  experiments}},\ }\href {https://doi.org/10.1103/PhysRevD.94.053007}
  {\bibfield  {journal} {\bibinfo  {journal} {Phys. Rev.}\ }\textbf {\bibinfo
  {volume} {D94}},\ \bibinfo {pages} {053007} (\bibinfo {year}
  {2016}{\natexlab{b}})},\ \Eprint {https://arxiv.org/abs/1606.05662}
  {arXiv:1606.05662 [hep-ph]} \BibitemShut {NoStop}%
\bibitem [{\citenamefont {Majhi}\ \emph {et~al.}(2020)\citenamefont {Majhi},
  \citenamefont {Chembra},\ and\ \citenamefont {Mohanta}}]{Majhi:2019tfi}%
  \BibitemOpen
  \bibfield  {author} {\bibinfo {author} {\bibfnamefont {R.}~\bibnamefont
  {Majhi}}, \bibinfo {author} {\bibfnamefont {S.}~\bibnamefont {Chembra}},\
  and\ \bibinfo {author} {\bibfnamefont {R.}~\bibnamefont {Mohanta}},\
  }\bibfield  {title} {\bibinfo {title} {{Exploring the effect of Lorentz
  invariance violation with the currently running long-baseline experiments}},\
  }\href {https://doi.org/10.1140/epjc/s10052-020-7963-1} {\bibfield  {journal}
  {\bibinfo  {journal} {Eur. Phys. J. C}\ }\textbf {\bibinfo {volume} {80}},\
  \bibinfo {pages} {364} (\bibinfo {year} {2020})},\ \Eprint
  {https://arxiv.org/abs/1907.09145} {arXiv:1907.09145 [hep-ph]} \BibitemShut
  {NoStop}%
\bibitem [{\citenamefont {Ayres}\ \emph {et~al.}(2007)\citenamefont {Ayres}
  \emph {et~al.}}]{NOvA:2007rmc}%
  \BibitemOpen
  \bibfield  {author} {\bibinfo {author} {\bibfnamefont {D.~S.}\ \bibnamefont
  {Ayres}} \emph {et~al.} (\bibinfo {collaboration} {NOvA}),\ }\bibfield
  {title} {\bibinfo {title} {{The NOvA Technical Design Report}},\ }\bibfield
  {journal} {\bibinfo  {journal} {FERMILAB-DESIGN-2007-01}\ }\href
  {https://doi.org/10.2172/935497} {10.2172/935497} (\bibinfo {year}
  {2007})\BibitemShut {NoStop}%
\bibitem [{\citenamefont {Huber}\ \emph {et~al.}(2005)\citenamefont {Huber},
  \citenamefont {Lindner},\ and\ \citenamefont {Winter}}]{Huber:2004ka}%
  \BibitemOpen
  \bibfield  {author} {\bibinfo {author} {\bibfnamefont {P.}~\bibnamefont
  {Huber}}, \bibinfo {author} {\bibfnamefont {M.}~\bibnamefont {Lindner}},\
  and\ \bibinfo {author} {\bibfnamefont {W.}~\bibnamefont {Winter}},\
  }\bibfield  {title} {\bibinfo {title} {{Simulation of long-baseline neutrino
  oscillation experiments with GLoBES (General Long Baseline Experiment
  Simulator)}},\ }\href {https://doi.org/10.1016/j.cpc.2005.01.003} {\bibfield
  {journal} {\bibinfo  {journal} {Comput. Phys. Commun.}\ }\textbf {\bibinfo
  {volume} {167}},\ \bibinfo {pages} {195} (\bibinfo {year} {2005})},\ \Eprint
  {https://arxiv.org/abs/hep-ph/0407333} {arXiv:hep-ph/0407333} \BibitemShut
  {NoStop}%
\bibitem [{\citenamefont {Huber}\ \emph {et~al.}(2007)\citenamefont {Huber},
  \citenamefont {Kopp}, \citenamefont {Lindner}, \citenamefont {Rolinec},\ and\
  \citenamefont {Winter}}]{Huber:2007ji}%
  \BibitemOpen
  \bibfield  {author} {\bibinfo {author} {\bibfnamefont {P.}~\bibnamefont
  {Huber}}, \bibinfo {author} {\bibfnamefont {J.}~\bibnamefont {Kopp}},
  \bibinfo {author} {\bibfnamefont {M.}~\bibnamefont {Lindner}}, \bibinfo
  {author} {\bibfnamefont {M.}~\bibnamefont {Rolinec}},\ and\ \bibinfo {author}
  {\bibfnamefont {W.}~\bibnamefont {Winter}},\ }\bibfield  {title} {\bibinfo
  {title} {{New features in the simulation of neutrino oscillation experiments
  with GLoBES 3.0: General Long Baseline Experiment Simulator}},\ }\href
  {https://doi.org/10.1016/j.cpc.2007.05.004} {\bibfield  {journal} {\bibinfo
  {journal} {Comput. Phys. Commun.}\ }\textbf {\bibinfo {volume} {177}},\
  \bibinfo {pages} {432} (\bibinfo {year} {2007})},\ \Eprint
  {https://arxiv.org/abs/hep-ph/0701187} {arXiv:hep-ph/0701187} \BibitemShut
  {NoStop}%
\bibitem [{\citenamefont {Denton}\ and\ \citenamefont
  {Gehrlein}(2023)}]{denton2023comes}%
  \BibitemOpen
  \bibfield  {author} {\bibinfo {author} {\bibfnamefont {P.~B.}\ \bibnamefont
  {Denton}}\ and\ \bibinfo {author} {\bibfnamefont {J.}~\bibnamefont
  {Gehrlein}},\ }\href@noop {} {\bibinfo {title} {Here comes the sun: Solar
  parameters in long-baseline accelerator neutrino oscillations}} (\bibinfo
  {year} {2023}),\ \Eprint {https://arxiv.org/abs/2302.08513} {arXiv:2302.08513
  [hep-ph]} \BibitemShut {NoStop}%
\bibitem [{\citenamefont {Esteban}\ \emph {et~al.}(2020)\citenamefont
  {Esteban}, \citenamefont {Gonzalez-Garcia}, \citenamefont {Maltoni},
  \citenamefont {Schwetz},\ and\ \citenamefont {Zhou}}]{Esteban:2020cvm}%
  \BibitemOpen
  \bibfield  {author} {\bibinfo {author} {\bibfnamefont {I.}~\bibnamefont
  {Esteban}}, \bibinfo {author} {\bibfnamefont {M.~C.}\ \bibnamefont
  {Gonzalez-Garcia}}, \bibinfo {author} {\bibfnamefont {M.}~\bibnamefont
  {Maltoni}}, \bibinfo {author} {\bibfnamefont {T.}~\bibnamefont {Schwetz}},\
  and\ \bibinfo {author} {\bibfnamefont {A.}~\bibnamefont {Zhou}},\ }\bibfield
  {title} {\bibinfo {title} {{The fate of hints: updated global analysis of
  three-flavor neutrino oscillations}},\ }\href
  {https://doi.org/10.1007/JHEP09(2020)178} {\bibfield  {journal} {\bibinfo
  {journal} {JHEP}\ }\textbf {\bibinfo {volume} {09}},\ \bibinfo {pages}
  {178}},\ \Eprint {https://arxiv.org/abs/2007.14792} {arXiv:2007.14792
  [hep-ph]} \BibitemShut {NoStop}%
\bibitem [{\citenamefont {Baker}\ and\ \citenamefont
  {Cousins}(1984)}]{Baker:1983tu}%
  \BibitemOpen
  \bibfield  {author} {\bibinfo {author} {\bibfnamefont {S.}~\bibnamefont
  {Baker}}\ and\ \bibinfo {author} {\bibfnamefont {R.~D.}\ \bibnamefont
  {Cousins}},\ }\bibfield  {title} {\bibinfo {title} {{Clarification of the Use
  of Chi Square and Likelihood Functions in Fits to Histograms}},\ }\href
  {https://doi.org/10.1016/0167-5087(84)90016-4} {\bibfield  {journal}
  {\bibinfo  {journal} {Nucl. Instrum. Meth.}\ }\textbf {\bibinfo {volume}
  {221}},\ \bibinfo {pages} {437} (\bibinfo {year} {1984})}\BibitemShut
  {NoStop}%
\end{thebibliography}%
\end{document}